\documentclass[twocolumn,dvipsnames]{aastex62}
\usepackage{amsmath,amstext,amssymb,mathtools,graphicx,color,xcolor}
\usepackage[T1]{fontenc}
\usepackage{apjfonts}

\usepackage[utf8]{inputenc}
\usepackage[figure,figure*]{hypcap}
\usepackage{chngpage}
\usepackage{url}
\usepackage{amsmath}   
\usepackage{soul}
\usepackage{array}
\usepackage[makeroom]{cancel}
\usepackage[normalem]{ulem}

\graphicspath{{./}{plots/}}

\newcommand{\edt}[2]{{#2}}

\submitjournal{ApJS}
\begin{document}

\shorttitle{Dynamic Relaxation for SPH Initial Data}    

\shortauthors{Kaltenborn, M.~A.~R. et al.}

\title{Halted-Pendulum Relaxation: Application to White Dwarf Binary Initial Data}  



%
\correspondingauthor{M. Alexander R. Kaltenborn}
\email{markaltenborn@lanl.gov}

\author[0000-0002-9604-7908]{M. Alexander R. Kaltenborn}
\affiliation{Center for Theoretical Astrophysics, Los Alamos National Laboratory, Los Alamos, NM, 87545, USA}
\affiliation{Theoretical Division, Los Alamos National Laboratory, Los Alamos, NM, 87545, USA}

\author[0000-0002-4510-7325]{Michael J. Falato}
\affiliation{Center for Theoretical Astrophysics, Los Alamos National Laboratory, Los Alamos, NM, 87545, USA}
\affiliation{Computer, Computational, and Statistical Sciences Division, Los Alamos National Laboratory, Los Alamos, NM, 87545, USA}

\author[0000-0003-4156-5342]{Oleg Korobkin}
\affiliation{Center for Theoretical Astrophysics, Los Alamos National Laboratory, Los Alamos, NM, 87545, USA}
\affiliation{Theoretical Division, Los Alamos National Laboratory, Los Alamos, NM, 87545, USA}

\author[0000-0002-1487-0360]{Irina Sagert}
\affiliation{Center for Theoretical Astrophysics, Los Alamos National Laboratory, Los Alamos, NM, 87545, USA}
\affiliation{Computer, Computational, and Statistical Sciences Division, Los Alamos National Laboratory, Los Alamos, NM, 87545, USA}

\author[0000-0002-5412-3618]{Wesley P. Even}
\affiliation{Center for Theoretical Astrophysics, Los Alamos National Laboratory, Los Alamos, NM, 87545, USA}
\affiliation{Theoretical Division, Los Alamos National Laboratory, Los Alamos, NM, 87545, USA}





\begin{abstract}
Studying compact star binaries and their mergers is integral to \edt{modern astrophysics}{determining progenitors for observable transients}. 
\edt{In particular, binary white dwarfs are associated with Type Ia supernovae, used as standard candles to measure the expansion of the Universe. }{}
Today, compact-star mergers are typically studied via state-of-the-art computational fluid dynamics codes. 
One such numerical technique\edt{s}{}, Smoothed Particle Hydrodynamics (SPH), is frequently chosen for its excellent mass, energy, and momentum conservation. 
\edt{Furthermore, t}{T}he natural treatment of vacuum and the ability to represent highly irregular morphologies make SPH an excellent tool for the \edt{numerical }{}study of compact-star binaries and mergers. 
\edt{However, f}{F}or many scenarios, including binary systems, the \edt{outcome simulations are }{outcome of simulations is} only as accurate as the initial conditions. 
For SPH, it is essential to ensure \edt{that particles}{that the particles} are distributed \edt{semi-regularly, correctly representing the initial density profile}{regularly, representing the initial density profile but without long-range correlations}. 
\edt{Additionally, p}{P}article noise in the form of high-frequency local motion and low-frequency global dynamics must be damped out. 
Damping the latter can be as computationally intensive as the actual simulation. 
\edt{Here, w}{W}e discuss a new and straightforward relaxation method, Halted-Pendulum Relaxation (HPR), to remove \edt{the }{}global oscillation modes of SPH particle configurations. 
In combination with effective external potentials representing gravitational and orbital forces, we show that HPR has an excellent performance in efficiently relaxing SPH particles to the desired density distribution and removing global oscillation modes. 
We compare the method to frequently used relaxation approaches \edt{such as gravitational glass, increased artificial viscosity, and Weighted Voronoi Tesselations, }{}and test it on a white dwarf binary model at its Roche lobe overflow limit.
\edt{}{We highlight the importance of our method in achieving accurate initial conditions and its effect on achieving circular orbits and realistic accretion rates when compared with other general relaxation methods.}
\end{abstract}


\section{Introduction}
\label{sec:intro}
\edt{Type Ia supernova (SN) explosions have been used in cosmology to provide the first evidence for the accelerated expansion of the Universe 
(Perlmutter et al. 1999) 
The general idea for the explosion mechanism is that a runaway fusion reaction of carbon and oxygen in the interior of a white dwarf (WD) produces so much thermonuclear energy that it completely disrupts the star 
(Hoyle \& Fowler 1960).
In the so-called single degenerate scenario, the white dwarf is in a binary with a giant branch or main-sequence star, accreting matter from its companion
(Canal et al. 1996).
As the mass of the white dwarf approaches the Chandrasekhar limit, temperature and pressure in its interior rise, leading to the ignition of fusion reactions.}{}
In a double degenerate scenario\edt{, the initial system is a WD binary (WDB) that merges and causes the supernova (Hoyle1960, Yungelson2017).}{ of Type Ia supernovae, a merger of a white dwarf binary (WDB) ignites thermonuclear supernova \citep{Hoyle1960, Yungelson2017}.}
The resulting observed light curves are expected to have \edt{peak luminosities consistent among Type Ia explosions}{highly uniform properties across all Type Ia explosions}, allowing their usage \edt{}{as standard candles} for distance determinations. 
However, at present, many open questions remain. 
In order to refine the distance estimate and, thereby, cosmological parameters, it is essential to understand how the properties of SNe Ia correlate with the age and type of their host galaxies \citep{Shariff2016}.
\edt{One example of such a correlation is the bias of more luminous Type Ia SNe originating from more massive galaxies (Sullivan2006).
In general, the so-called ``standardization parameters'' used for calibrating Type Ia light curves are dependent on redshift, reflecting not only observational bias but also the evolution of the progenitors over cosmic history.
A fundamental problem within supernova cosmology is that fiducial models describe these dependencies.
However, such approaches lack an understanding of the underlying evolution of the progenitors.
For example, 
(Childress2013) 
revealed an unexplained 4$\sigma$ preference for the ``z-jump color correction model'' for high-redshift supernovae, which could be caused by a shift in the progenitor type at a particular moment in the cosmic time.
Furthermore,}{For instance,} observations and population synthesis both predict insufficient formation rates of near-Chandrasekhar WDs to explain the observed rates of SNe Ia. 
As a consequence, sub-Chandrasekhar mergers must also contribute as SNe Ia producers \citep{vanKerkwijk2010}.

Resolution of these problems requires more accurate numerical modeling of the double degenerate scenario \citep{Fryer2008}. 
Historically, the first models to simulate white dwarf mergers used Smoothed Particle Hydrodynamics (SPH) \citep{Benz1990, Rasio1995, Segretain1997, Guerrero2004, Yoon2007}.
Subsequent improved SPH techniques provided extensive simulations of the double degenerate scenario and its various aspects \citep{Dan09, Fryer2010, Dan11, Dan12, Raskin2012, Raskin2014, Dan2014, Rosswog2015}.
However, quirks of early SPH approaches prompted researchers to develop alternative methods based on the grid representation in the co-rotating frame, with particular attention to angular momentum conservation \citep{Motl2002, DSouza2006, Motl2007}. These methods were developed alongside capabilities to generate highly accurate initial conditions for contact binaries using self-consistent field methods \citep{Hachisu1986a, Hachisu1986b, Even2009}.   
The qualitative discrepancies in the outcomes between the two approaches raised concerns about the accuracy of the treatment of the problem, especially regarding the initial data \citep{Dan09, Fryer2008, Zhu2013}.
To address the discrepancies between SPH and grid-based methods, \cite{Motl2017} conducted a thorough comparison study and found similar merger outcomes for various models.
Other numerical techniques were proposed, such as a moving mesh \citep[with the code AREPO,][]{Pakmor2016}, curvilinear co-rotating mesh \citep{Kadam2018}, and adaptive mesh \citep{Katz2016}.

\edt{}{Another aspect of merger simulations is an accuracy of internal flows within the components of the binary. 
The quality of these simulations can be compromised if these internal flows are driven by numerical errors.
For example, the behavior of a layer of potential nuclear fuel in a direct impact system cannot be addressed if passive scalars (in the Eulerian case) or stratified particle layers are unphysically mixed \citep{Dan11}.}

An improved numerical methodology can provide a more in-depth understanding of the double degenerate scenario.
Simulations of WDBs demand that an intricate balance of energy and angular momentum be maintained; in particular, improved conservation of angular momentum is essential \citep{Marcello2017, Korobkin2021, Marcello2021}.
While achieving conservation is easier in the context of SPH, this method presents its unique problem: the configuration of particles must accurately represent the initial density distribution.
In the standard approach, sometimes called ``gravitational glass,'' the system is set up to relax to the state with zero velocities. 
The particles are allowed to move slowly, with a damping term proportional to particle velocities, until they find an equilibrium supported by pressure, Newtonian forces, and an optional effective centrifugal potential to account for rotation. 
Other proposed methods include Weighted Voronoi Tesselations \citep[WVT,][]{Diehl2015} and the Adaptive Pressure Method \citep[APM,][]{Rosswog2020}.

However, all of these relaxation techniques present a problem. 
While high-frequency local motion is quickly damped, the global oscillation modes with lower frequency take much longer to suppress. 
Here, we present and test a straightforward idea for particle relaxation, which has been overlooked so far. 
The process of velocity damping for low-frequency modes is akin to the damped pendulum oscillations, where the kinetic energy of the pendulum is periodically converted to the potential energy of particle pressure and vice versa. 
We propose monitoring the system's total kinetic energy and halting the particles at the point where the system reaches its maximum. 
Since the particle configuration's total energy (kinetic $+$ potential) is conserved, this simple trick naturally finds a minimum of potential energy.

In this paper, we focus on setting up initial conditions in WDBs. 
We first test our method on single stars and demonstrate that the global breathing mode is completely suppressed after only a few ``pendulum halts.''
Moreover, the total kinetic energy of spurious particle motion is lowered by a few orders of magnitude. 
We also apply our method to the initial configuration for a double WD system that will undergo mass transfer and evolve the configuration for several orbits.
In addition to resetting particle velocities via our relaxation approach, we also apply effective potentials to determine particle accelerations from gravity and the orbital motion in a binary.
Notice that SPH particle relaxation in a self-consistent gravitational potential essentially solves an elliptic equation for the stellar structure.
This implies that the results below can be extended to other problems where elliptic equations \edt{}{for stable particle equilibrium} must be solved, including quasistationary sequences of binary systems and solitary stars with uniform or differential rotation~\citep{Yoshida2019}.

This paper is organized as follows: 
Section \ref{sec:sph} briefly introduces SPH and the FleCSPH code. 
Section \ref{sec:method} outlines the numerical methods for setting up initial configurations in various systems. 
Section \ref{sec:results} describes the results of numerical evolution for the generated particle configurations for various systems, including single and binary WDs. 
We summarize our findings in Section~\ref{sec:conclusion}. 
\section{Brief Overview of SPH}
\label{sec:sph}
SPH is a fully Lagrangian mesh-free numerical method that solves the hydrodynamic equations by discretizing the flow into fluid elements called particles. 
The method is founded upon the notion of interpolation. 
The conservation laws for hydrodynamic flows are transformed into integrable equations using an interpolation function called the \textit{kernel}. 
The flow characteristics at particle $a$'s location are evaluated by averaging particle properties in the neighborhood $\Omega$ of $a$. 
The average sum is computed using the kernel $W$ as a weight. 
With that, a quantity $A$ at the location $\vec{r}_a$ is given by

\begin{align}\label{eq:sph_fundamental}
    A_a = A(\vec{r}_a) \simeq \sum_{b \: \in \: \Omega}V_b \: A_b \: W(\left|\vec{r}_{ab}\right|, h) \text{,} \:\:\: \vec{r}_{ab} = \vec{r}_a-\vec{r}_b,
\end{align}
where $h$ is the so-called smoothing length (or hydrodynamic interaction range), and $V_b$ is a volume element.
The kernel function $W(\left|\vec{r}_{ab}\right|,h)$ is often abbreviated as $W_{ab}$. 
It is typically a function with compact support that reduces to the delta function in the limit of vanishing $h$. 
The volume element is determined by 

\begin{align}\label{eq:vol_ell}
    V_b = m_b/\rho_b, 
\end{align}
although other options are possible \citep{Saitoh2013, Hopkins2013, Rosswog2015}.
Using Eqs.(\ref{eq:sph_fundamental}) and (\ref{eq:vol_ell}), the density at $\vec{r}_a$ can simply be written as 

\begin{align}\label{eq:sphrho}
    \rho_a = \sum_b m_b \: W_{ab} \: \text{.} 
\end{align}

In SPH, spatial derivatives also move from operating on the physical quantities to the interpolation kernel. 
The derivative of the function $A$ is determined by differentiating the discretized formulation, i.e.,

\begin{align}\label{eq:sphder}
    \nabla\cdot{A}_a &= \nabla \cdot \sum_b \frac{m_b}{\rho_b} \: A_b \: W_{ab} = \sum_b \frac{m_b}{\rho_b} \: A_b \: \nabla \cdot W_{ab} \text{,} 
\end{align}
where the divergence in the sum operates with respect to the primed, or $a^{\textrm{th}}$, coordinate.
With that, the SPH equations for the conservation of linear momentum and energy can be written as 

\begin{align}
    \frac{d \vec{v}_a}{d t} &= -\sum_b m_b \left( \frac{P_a}{\rho_a^2} + \frac{P_b}{\rho_b^2}+ \Pi_{ab} \right) \nabla_a W_{ab}+ \vec{g}_a\text{,}\label{eq:sphp}\\
    \frac{d u_a}{dt} &= \sum_b m_b\left(\frac{P_a}{\rho_a^2} + \frac12\Pi_{ab}\right)\vec{v}_{ab} \cdot \nabla_a W_{ab}\text{,}\label{eq:sphe}
\end{align}
where $\Pi_{ab}$ is the viscous stress tensor which can be defined in a few different ways, as we will briefly discuss later in the paper.
Equations (\ref{eq:sphrho}), (\ref{eq:sphp}), and (\ref{eq:sphe}) are the system of hydrodynamic equations that are solved in FleCSPH in combination with a problem-dependent equation of state (EoS).

SPH is well-suited for physical scenarios like compact binary mergers (CBMs), as it can handle complex geometries and deformations and naturally supports true vacuum conditions.
The method also has intrinsic conservative properties, i.e., the conservation of mass is included by construction, and the conservation of linear momentum, angular momentum, and energy can be implemented up to machine precision.
SPH has been widely used since 1977 when Gingold and Monaghan published the first work on SPH to simulate non-spherical oscillations of stars \citep{Gingold1977}. 
Today, SPH in astrophysics is used for simulations of asteroidal and planetary collisions \citep{Maindl2013, Kegerreis2018}, core-collapse supernovae \citep{Ellinger2012}, and neutron-star mergers \citep{Rosswog1999}. 
Some examples of modern SPH codes for astrophysics are MAGMA2 \citep{Rosswog2020}, SPHINCS\_BSSN~\citep{Rosswog2021}, SPHERAL~\citep{spheral}, Phantom~\citep{Price2018}, miluphcuda~\citep{Schaefer2019}, and SNSPH~\citep{Fryer2006} codes.
Here, we use FleCSPH \citep{Loiseau2020a}, an open-source distributed SPH code, which is built on top of the Flexible Computational Science Infrastructure (FleCSI) developed at Los Alamos National Laboratory (LANL) \citep{charest2017flexible}. 
Details of the code can be found in, e.g., \cite{Loiseau2020b}. 
It contains analytic and tabular astrophysical equations of state and gravitational force calculation via the Fast Multipole Method. 
FleCSPH is applied to the simulation of compact star mergers \citep{Loiseau2020b} and their ejecta \citep{Stewart2022} as well as neutron star crust dynamics \citep{Sagert2022, Tsao2021}.
Here, we discuss white dwarf (WD) binary systems with a focus on their initialization.
\section{SPH Particle Initialization and Relaxation} 
\label{sec:method}
As previously mentioned, the initial particle placement in SPH can be crucial to ensure the correct dynamical behavior of a simulation and, therefore, a meaningful final result.
If the density variations in the physical system are small compared to the density itself, a regular lattice with varying particle mass can accurately reproduce initial conditions. 
However, when density gradients in the problem are so strong that the resulting particles have very different masses, SPH simulations tend to fail with noisy particle motion \citep{Lombardi1999}. 
Equal mass particles are preferred for modeling such problems.
In this case, either the particle lattice must be stretched and squeezed to fit the desired density distribution \citep{Price2018}, or the particles must be cleverly placed using other geometric or physically-informed methods.
Two examples of such methods for particle initialization in SPH are the Weighted Voronoi Tesselations~\citep[WVT,][]{Diehl2015} and Artificial Pressure Method~\citep[APM,][]{Rosswog2020}.
As discussed in this paper, these methods may still retain a small amount of global motion, which can become significant for accurate binary simulations.
For stars, one example of such motion is the so-called breathing mode---a large-scale collective oscillation of particle densities and positions. 
A method like WVT or gravitational glass operates locally and, therefore, is not capable of removing the low-frequency global mode.
The \emph{Halted-Pendulum Relaxation} (HPR) technique we introduce here is specifically targeting such low-frequency modes and can be used in conjunction with other methods for best results.
\subsection{Initialization}
For a \edt{single}{spherically-symmetric, non-rotating} star in hydrostatic equilibrium, the stellar structure can be obtained by solving 
\edt{Lane-Emden-type equation [Lane1870], which is a form of the Poisson's equation for the gravitational potential of a Newtonian, self-gravitating, spherically-symmetric fluid.}{an elliptic Poisson equation, which reduces to Lane-Emden equation for the case of a polytrope \citep{Lane1870,Emden1907}.} 
It is straightforward to obtain accurate solutions for the 1D radial density and pressure profiles.
With the profiles in hand, the SPH particle positions can be assigned using, e.g., the icosahedral shell method of \cite{Tegmark96}. 
This approach works well for spherical surfaces and allows mapping to the density profile while imposing equal particle mass across a desired number of particles; see Figure~\ref{fig:1}. 
The result of the icosahedral particle placement is a highly regular lattice of initialized data which, however, can seed undesired numerical effects during the subsequent dynamical evolution of the star. 
To avoid that, we add a small perturbation to the lattice placement, using a perturbation magnitude proportional to a small fraction of the particle's smoothing length with a random angular direction.
We then allow the particles to relax to a pseudo-ordered distribution using different methods described in the following subsections. 
\subsection{Energy Dissipation via Numerical Viscosity}
In this method, viscous particle interactions are raised artificially high to aggressively dissipate the kinetic energy of global modes and particle noise.
In Eq.~(\ref{eq:sphp}), we modify the momentum conservation by adding the numerical viscosity tensor $\Pi_{ab}$:

\begin{align}\label{eq:mg_visc}
    \Pi_{ab} = \begin{cases}
  \frac{\alpha \bar{c}_{ab} \mu_{ab} + \beta \mu^2_{ab}}{\bar{\rho}_{ab}}  & \text{for}\:\:\:\vec{r}_{ab}\cdot\vec{v}_{ab} < 0, \\
  0 & \text{otherwise,}
\end{cases} 
\end{align}
where 

\begin{align}
    \mu_{ab} &= \frac{\bar{h}_{ab}\vec{r}_{ab}\cdot \vec{v}_{ab}}{r^2_{ab}+\epsilon \bar{h}_{ab}^2} \text{,} \:\:\:
    r_{ab} = \left|\mathbf{r}_{ab}\right| = \left|\mathbf{r}_a - \mathbf{r}_b\right| \text{,} \nonumber\\ 
    \bar{h}_{ab} &= (h_a + h_b)/2 \text{,} 
\end{align}
$\bar{c}_{ab}$ is the average sound speed of particles $a$ and $b$, and $\bar{\rho}_{ab}$ is their average density. 
The parameters $\alpha$ and $\beta$ determine the strength of the dissipation and are typically set to $\alpha \sim 1$ and $\beta = 2 \alpha$, which leads to good agreement with, e.g., shock benchmark tests \citep{Monaghan1992}.
Increasing the values for both parameters can significantly increase the dissipation of the kinetic energy. 
With that, we can set viscosity parameters to be higher than their typical values (we chose $\alpha = 5$) and run the regular dynamical evolution of the star. 
One advantage of this method is that it is trivial to apply. 
Most SPH codes already contain the widely used numerical viscosity term of Eq.~(\ref{eq:mg_visc}), and the relaxation step can be run by simply raising the viscosity values.
However, caution must also be applied when using this method.
If the kinetic energy is dissipated too aggressively, the particles can stop moving before reaching a relaxed state. 
The breathing mode will reappear again when using such an initial state in a simulation with normal viscosity. 
Tests and adjustments for the viscosity parameters are\edt{ most likely always}{ }required to find values for efficient yet accurate relaxation.
\subsection{WVT Relaxation}
This relaxation approach uses repulsive interactions between particles within a smoothing length of one another.
The particle displacement due to the forces is guided by the particles' smoothing lengths, which are determined from the target density profile.

\cite{Diehl2015} introduced this method with a force that depends on the inverse of the squared inter-particle distance $r_{ab}$.
At each iteration, the displacement of particle $a$ is given by:

\begin{align}
\Delta \vec{r}_a &= \mu_\mathrm{wvt} \: \vec{D}_a, \\
\vec{D}_a &= h_a \: \sum_b \left[\left(\frac{\bar{h}_{ab}}{r_{ab} + \epsilon}\right)^2 + \kappa \right] \frac{\vec{r}_{ab}}{r_{ab}}, \\
\kappa &= - \left(\frac{\bar{h}_{ab}}{\bar{h}_{ab} + \epsilon}\right)^2. 
\label{diehl_deltax}
\end{align}

The factor $\mu_\mathrm{wvt}$ specifies the fraction of $h_a$ that particle $a$ can move during each iteration step. 
It is recommended to decrease $\mu_\mathrm{wvt}$ during the relaxation process.  
This allows large position displacements initially and restricts the particle motion as the relaxed state approaches. 
The value of $\epsilon$ can be set to a small fraction of the smoothing length to avoid numerical problems for particles that are close together.
Finally, the constant $\kappa$ is added to the force to ensure that the latter vanishes for $r_{ab} > \bar{h}_{ab}$.

The WVT-like method introduced by \cite{Arth2019} is very similar. 
The most significant difference lies in the repulsive force, which uses the SPH kernel. 
At each iteration, the particle displacement is calculated by 

\begin{align}
\Delta \vec{r}_a = \mu_\mathrm{wvt} \: \vec{A}_a, \:\:\: \vec{A}_a = \sum_b \bar{h}_{ab} \: W_{ab} \: \frac{\vec{r}_{ab}}{r_{ab}}\text{.}
\label{arth_deltax}
\end{align}

The value of the smoothing length is normalized in both methods according to the desired number of particle neighbors $N_\mathrm{ngb}$. 
In our implementation, at each iteration, we determine the target density $\rho_t$ for a particle's location $\rho_{t,a} = \rho_t (\vec{r}_a)$ and the corresponding smoothing length. 
Following \cite{Diehl2015}, we calculate the latter via 

\begin{align}
h_{t,a} = \eta \left(\frac{m}{\rho_{t,a}}\right)^{1/3}\text{,}
\label{h_diehl}
\end{align}
while \cite{Arth2019} suggests a different formulation that is tied to the desired number of neighbors:

\begin{align}
h_{t,a} = \left( \frac{3}{4} \frac{N_\mathrm{ngb} \: m}{\pi \rho_{t,a}} \right)^{1/3} \text{.}
\label{h_arth}
\end{align}

After the target smoothing lengths have been set, the sum of all individual SPH particle volumes is calculated via:

\begin{align}
V_\mathrm{SPH} = \frac{4}{3} \pi \: \sum_i h_{t,i}^3. 
\end{align}

The smoothing lengths are then scaled so that $V_\mathrm{total} = V_\mathrm{SPH} / N_\mathrm{ngb}$. 
With $V_\mathrm{total} = (4 \pi/3) \: r_\mathrm{star}^3$ where $r_\mathrm{star}$ is the radius of the star, the scaling is given by 

\begin{align}
h_a = \alpha \: h_{t,a}, \:\:\: \alpha = \left( \frac{N_\mathrm{ngb}}{\sum_i h_{t,i}^3} \right)^{1/3} r_\mathrm{star}\text{.}
\end{align}

To implement boundary conditions, \cite{Diehl2015} recommend ghost particles as these are commonly used in SPH codes. 
The ghost particles are created by mirroring simulation particles within a distance of one or two smoothing lengths from the boundary. 
\cite{Arth2019} apply periodic boundary conditions followed by cutting out the region of interest once convergence is reached.

In our implementation of this method in FleCSPH, we mirror the neighbors of each particle with $r_a+h_a \leq r_\mathrm{star}$, where $r_a$ is the radial distance of the particle from the star's center. 
A particle can interact with mirror images just like with real particles. 
Spherical mirroring at the sphere's radius is straightforward but leads to a slightly lower density of the mirrored particles. 
An option that guarantees the same density of the reflected particles is using a plane mirror that is normal to a particle's position vector. 
Such a mirror could be positioned either at the intersection of the particle position vector with the star or at the particle's location. 
Repulsive forces of particles and mirror particles parallel to a particle's position vector cancel out, and only contributions that are normal to the position vector remain. 
We implement this boundary condition by computing the total displacement for a particle due to its neighbors, removing the contribution parallel to the particle position vector, and doubling the normal one. 
To check for convergence, we follow \cite{Arth2019} and monitor the relaxation process, including a condition for its termination.
The authors suggest stopping the relaxation when most particles move less than a small fraction of the mean particle spacing. 
The latter is calculated via

\begin{align}
	d_{a} = N_\mathrm{ngb}^{-1/3} \: h_a. 
\end{align}

At each iteration, we count the number of particles $N_\mathrm{cnt}$ that experience a displacement $|\Delta \vec{r}_a | > 10^{-3} \: d_{a}$. 
When $N_\mathrm{cnt}$ is smaller than a user-specified percentage of $N_\mathrm{total}$, the relaxation is stopped.

The advantage of WVT-like relaxation is that the method is very fast. 
Although it does require additions to a regular SPH code, these are not very extensive, especially when the smoothing kernel is applied in the repulsive force. 
However, our WVT implementation leads to a final configuration that significantly deviates from the analytic solution for the star's density and pressure profile. 
Since the method does not directly compare the particle density to the target value during relaxation, assessing its success or failure is not straightforward. 
In FleCSPH, this is only done after the relaxed star has evolved, and the density is determined via the standard SPH techniques. 
However, given that the method has been successfully applied in other studies, future work is likely required to improve the WVT implementation in FleCSPH.
\subsection{Relaxation in External Potential}
\label{sec:extpotrel}
In this method, an artificial external potential is designed to provide a potential force pushing the particles to the correct equilibrium configuration.
To remove any spurious motion, particle velocities are damped with an additional drag term in acceleration, $\propto -\beta_{\rm drag}\vec{v}$.
For the case when internal self-gravity forces of the system are used instead of external potentials, this method is known as ``gravitational glass'' \citep{White1996, Wang2007, Diehl2015}, referring to the type of irregular particle lattices that it generates.

In general, external forces are routinely applied in computational fluid dynamics codes. 
The simplest example is the inclusion of gravitational acceleration in terrestrial problems, e.g., as applied in the simulation of Rayleigh-Taylor instabilities. 
In FleCSPH, we include external potentials and the resulting acceleration of particles ${\vec{a}_{\rm ext} = -\nabla\varphi_{\rm ext}}$\edt{ }{.}
Implemented potentials include spherical and parabolic potential walls, a 2D airfoil in a wind tunnel, artificial drag, and binary orbital potentials. 
\cite{Tsao2021} has discussed the usage of potential walls for the simulation of the neutron-star crust as a thin shell. 
Here, we will present two examples of external potentials for the single star relaxation and the relaxation of compact star binaries. 
\subsubsection{Single Star Relaxation}
The general idea for external potential relaxation is to obtain an equilibrium configuration, i.e., where the momentum equation, including an external acceleration, is

\begin{align}
    \frac{d\vec{v}}{dt} = - \frac{1}{\rho} \: \nabla P + \vec{a}_{\rm ext} = 0\text{,}
\end{align}
which directly results in

\begin{align}\label{eq:aext}
    \vec{a}_{\rm ext} = \frac{1}{\rho} \: \nabla P \text{.} 
\end{align}

For a spherical star with a given density and pressure profile, as, e.g., obtained by the Lane-Emden equation, one can determine $\nabla P (r) = dP(r)/dr$ and, therefore, directly calculate the necessary $\vec{a}_{\rm ext} (\vec{r})$. 
Furthermore, when using a polytropic EoS 

\begin{align}
    P(\rho) = K \rho^\Gamma \text{,}
\end{align}
with problem-specific parameters $K$ and $\Gamma$, the acceleration can be written as

\begin{align}
    \vec{a}_{\rm ext} (r) = K \: \Gamma \: \rho(r)^{\Gamma - 2} \: \frac{d\rho}{dr}\text{,}
\end{align}
while the potential is

\begin{align}
    \varphi_{\rm ext} (r) = - \frac{K \Gamma \rho(r)^{\Gamma - 1}}{\Gamma - 1}\text{.}
\end{align}

Thus, both only require information on the density profile. 
Another advantage of the polytropic EoS is that by setting $\Gamma \sim 1$, the sound speed,

\begin{align}
    c_s (r) = \sqrt{\Gamma \: K \: \rho(r)^{\Gamma -1}}\text{,}
\end{align}
is uniform throughout the star. 
Since the sound speed is often the decisive factor for the time step size in a simulation, this can vastly improve the numerical efficiency of the relaxation process, especially for stars with significant density variations.
Even though the polytropic EoS is used, the relaxation method is not restricted only to stars described by it. 
In practice, single stars can initially be relaxed using a polytropic EoS with $\Gamma \sim 1$. 
This results in the particle configuration which correctly reproduces the equilibrium density profile (although perhaps wrong values for the pressure and other thermodynamic quantities, which can be recomputed in a separate modification step).

%
%

\subsubsection{Roche-Potential Relaxation for Binary Systems}
To prepare a system of two stars (referred to as stars 1 and 2) in a binary, we relax the particles in the Roche potential, consisting of the self-gravity forces and the non-inertial forces in the co-rotating frame.
In FleCSPH, self-gravity interactions are computed with the Fast Multipole Method (FMM) method \citep{Dehnen2002}.
In the reference frame co-rotating with the binary at an angular velocity $\Omega$, 

\begin{align}
    \Omega = \sqrt{\frac{G(m_1 + m_2)}{a^3}}.
\end{align}

where $a$ is the orbital separation of the binary, 
the individual SPH particles are subject to centrifugal and Coriolis forces. 
The total angular momentum of a binary in a circular orbit with point masses $m_1$ and $m_2$ is given by:

\begin{align}\label{eq:13}
    \ell = m_1 m_2\sqrt{\frac{G a}{m_1 + m_2}}, 
\end{align}

In the non-inertial frame uniformly rotating with angular velocity $\Omega$ around the $z$-axis, the particles experience a total acceleration

\begin{align}
    \vec{a}_r = \vec{a}_i - 2\:\vec{\Omega}\times\vec{v} - \vec\Omega \times(\vec\Omega\times\vec{r}),
    \label{eq:inertia}
\end{align}
where $\vec{a}_i$ is the acceleration in an inertial frame, and the following two terms are the Coriolis and centrifugal accelerations, respectively.
For the remainder of the paper, we will focus on WDBs, which are tidally locked. Thus, we only need to implement the centrifugal terms to the particle accelerations and effective potentials:

\begin{align}
 \vec{a}_r &= \vec{a}_i  + \Omega^2\vec{R}, \\
 \varphi_r &= \varphi_i  - \frac12\Omega^2 R^2.
\end{align}
where $\vec{R}$ is the projection of the position vector $\vec{r}$ onto the $xy$-plane.

To enforce angular momentum conservation during the relaxation process, we recompute $\Omega$ every time step, using the expression

\begin{align}
 \Omega &= \frac{\ell}{\sum_{a} m_a R_a^2}.
 \label{eq:omega}
\end{align}
The initial angular velocity of the binary does not match the angular velocity of the two extended bodies. 
If a stable orbital configuration exists with the given angular momentum, the value of $\Omega$ will eventually adapt and stabilize the stars around the correct orbital separation. 
This method acts to deform the stars accordingly and gives another approach to more accurate initial conditions for binary orbits. 
Once the system is properly equilibrated to the stars' configurations in orbit, we fully evolve the system by mapping the orbital velocities onto each particle.
\subsection{Halted-Pendulum Relaxation}
While the usual methods for particle relaxation are local in that they target and rely upon local particle distributions, the HPR method is global because it relies on the total kinetic energy. 
It is motivated by the idea that stopping an oscillating pendulum at the point of its highest kinetic energy will leave it at the point with the lowest potential energy, which is precisely what we are trying to find. 
As described by the SPH Hamiltonian, a system of particles possesses the usual spectrum of oscillation eigenmodes near an equilibrium configuration (if such configuration exists). 
With the standard initialization techniques, the energy is distributed randomly between various eigenmodes. 
While local particle relaxation methods suppress high-oscillation frequencies, some energy remains in the lower modes because these modes tend to be the most global. 
The lowest fundamental mode usually has the size of the system. 
Considering the SPH system near the particle equilibrium with the minimum potential energy as an ensemble of oscillators, it is easy to see why draining the kinetic energy from the system when it reaches its peak will lead to it rapidly approaching the potential minimum.


When applying the HPR method to a single star or binary, the total kinetic energy of the system must be monitored. 
When total kinetic energy has reached a maximum, the velocities of all particles in the system are set to zero. 
The simulation is continued, with repeated HPR steps as necessary. 
On a technical level, the trend in the kinetic energy can be monitored by storing a few previous values and making a polynomial regression to determine if a maximum has been reached. 
When the maximum is found, the velocities of all particles are set to zero.

There is an elegant simplicity to the HPR method and remarkable effectiveness, as shown in the next section. 
In practice, we found that HPR only needs to be triggered on the order of ten times or fewer before the noise floor for a system is reached.
\section{Numerical Results}
\label{sec:results}
Below we present the description of the setup procedure, particle relaxation, and subsequent evolution of single and binary WD systems.
\subsection{Single stars}
\begin{figure*}
\begin{tabular}{cc}
  \includegraphics[width=0.47\textwidth]{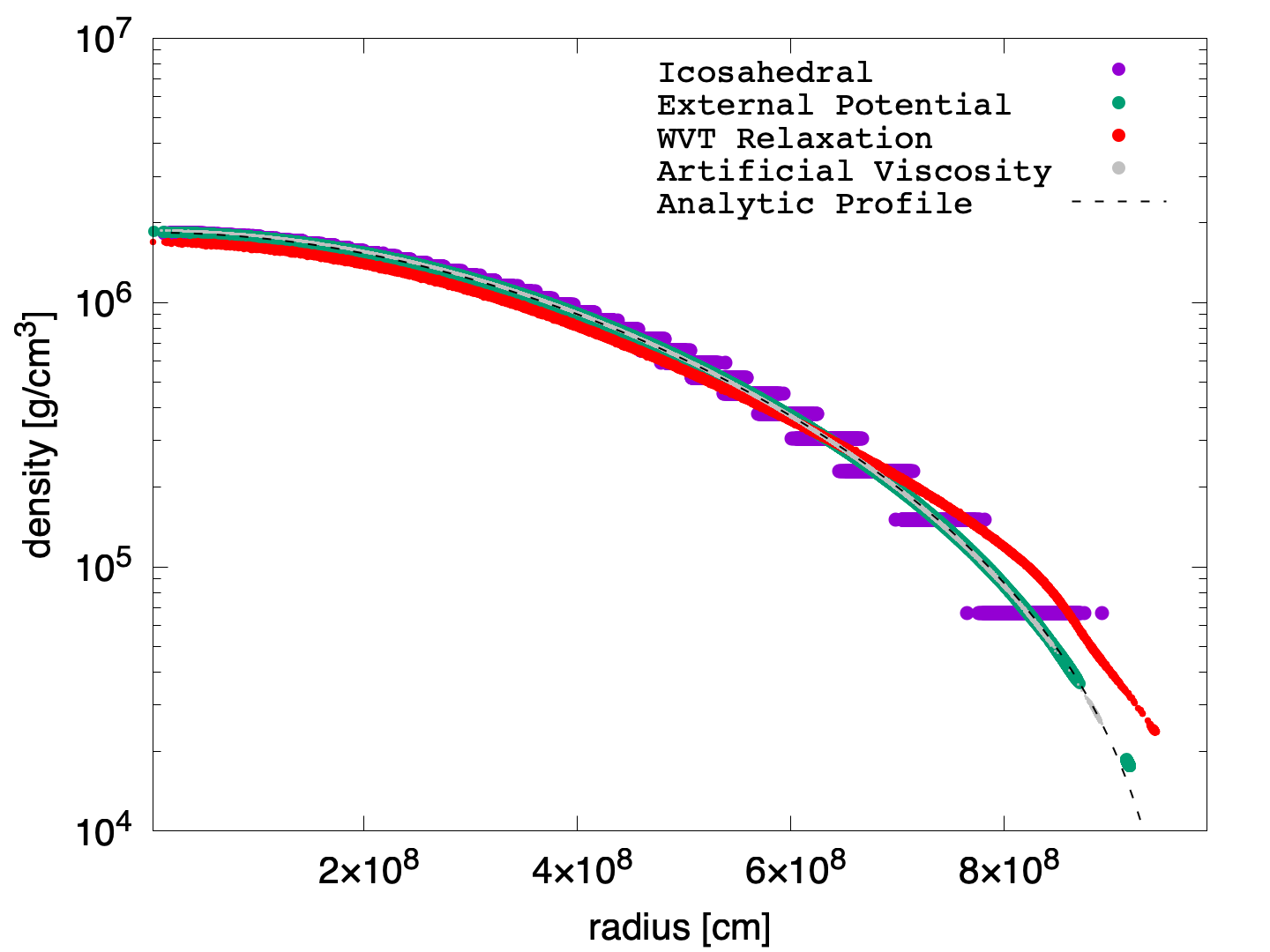} &
  \includegraphics[width=0.47\textwidth]{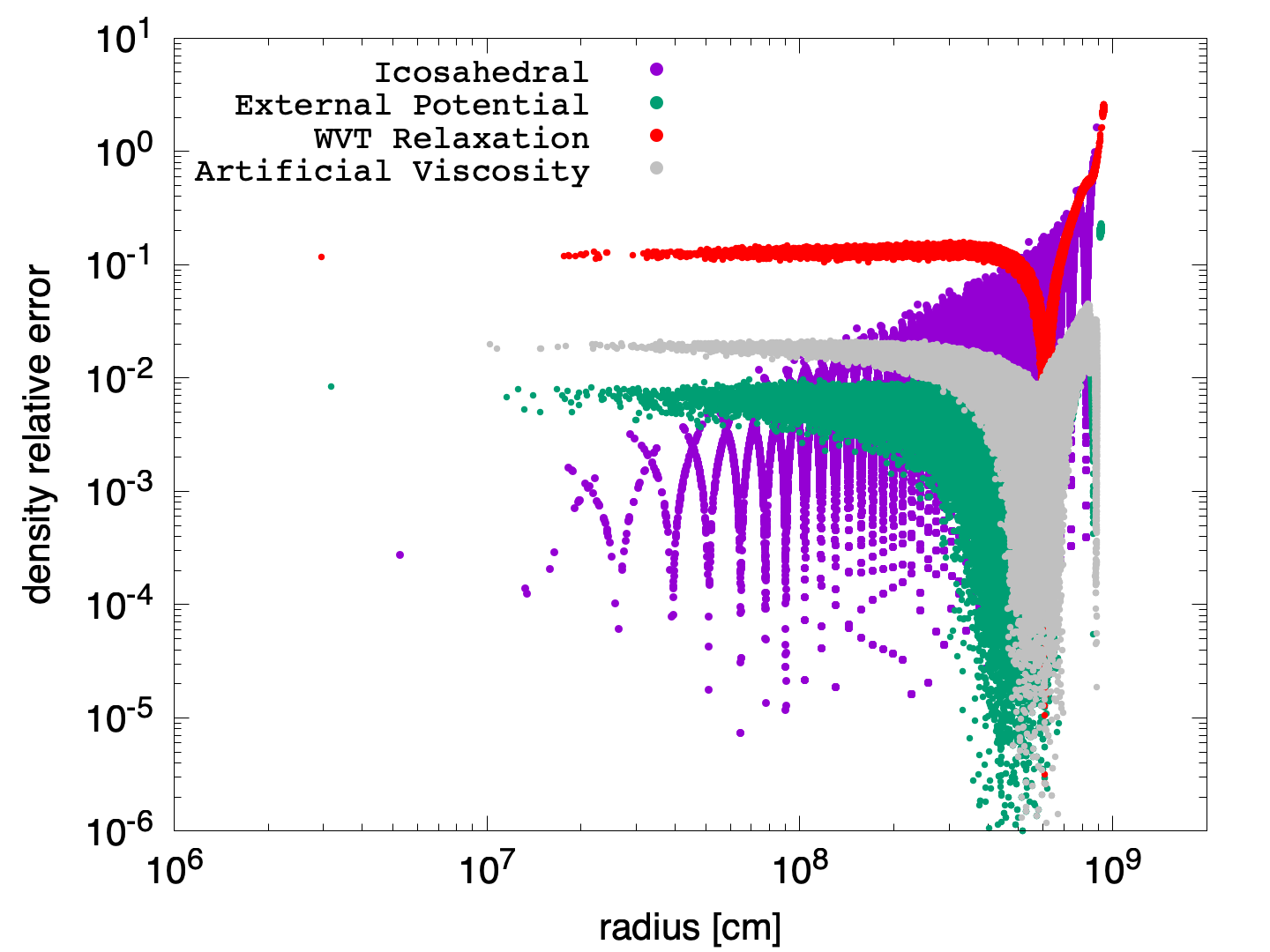}
\end{tabular}
\caption{Left: target radial density profile (dashed line) for a single WD in comparison to the density computed for various particle configurations representing this profile, from top to bottom: icosahedral shells with initial perturbation; velocity damping in external potential\edt{; density after WVT relaxation using WVT density estimator}{}; density after WVT relaxation using\edt{ regular}{} SPH density estimator; and increased artificial viscosity.
Right: the relative error between the target profile and the densities for various particle configurations.
\edt{}{
We find the external potential method provides the smallest error. The WVT relaxation relies on a geometric average distance between particles to relax the system into a semi-regular configuration. However, while this method produces highly regular configurations, when the configuration is handed over to the SPH solver, the density significantly departs from the desired profile.}
}
\label{fig:1}
\end{figure*}

We use a cold WD EoS to set up a single $0.5\ M_\odot$ star at three different resolutions using $5 \times 10^4$, $10^5$, and $2\times 10^5$ particles.
After setting up the particles via the icosahedral shells with small random positional perturbations, we apply three particle relaxation techniques: velocity damping in the external potential, WVT, and artificial-viscosity relaxation. 
Figure~\ref{fig:1} shows the radial profiles after initialization and for the final step of the three relaxation methods. 
It can be seen that the icosahedral distribution follows the target density profile but introduces step-like features which are caused by the random positional perturbations in each shell.
The external potential method produces a particle configuration that best fits the target profile with an accuracy of $10^{-2}$. 
The WVT relaxation is efficient and, within its density estimator, manages to fit the desired density profile much better, at an accuracy of $10^{-3}$ or less.
However, it disagrees considerably with the target density when this density is calculated with the regular SPH estimator (Eq.~\ref{eq:sphrho}).
This should not be regarded as a failure of WVT but rather as a point where more effort must be applied in FleCSPH to use the method. 
We take advantage of this result to demonstrate that HPR can be efficiently applied, even if the initial system needs to fit the desired profile better.

\begin{figure}[htp!]
\centering
 \includegraphics[width=\columnwidth]{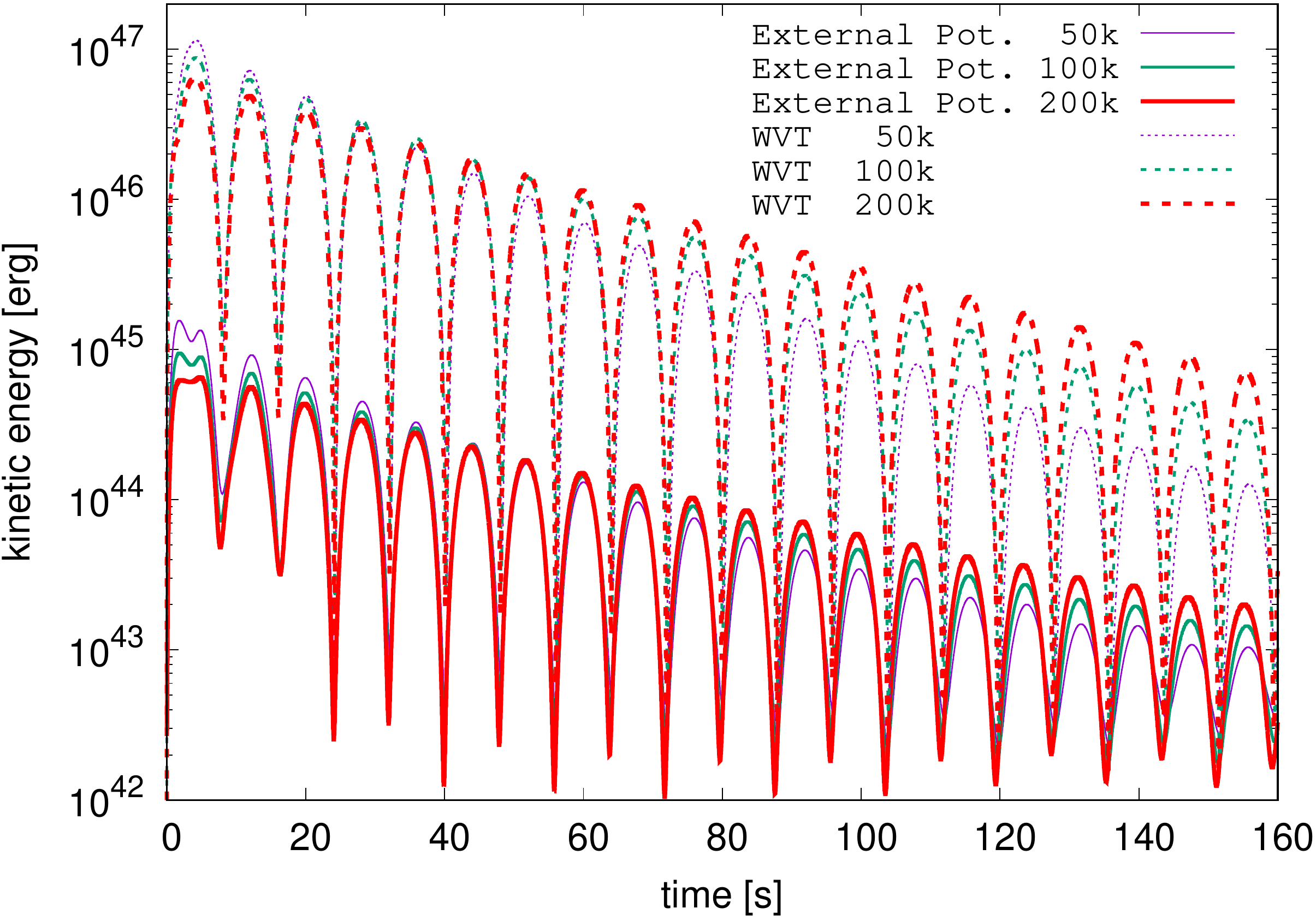}
\caption{Evolution of the kinetic energy for the star relaxed with two methods: WVT (dashed lines) and velocity damping in an external potential (solid lines), for three different resolutions. 
The breathing mode for the star relaxed with WVT has a much larger amplitude than for the one relaxed with the effective potential method.
Viscous damping is less efficient for higher resolutions.}
\label{fig:global-breathing-mode}
\end{figure}

Even after the relaxation, all particle configurations possess global modes that we seek to remove.
Figure~\ref{fig:global-breathing-mode} shows the free hydrodynamic evolution of the total kinetic energy of the star relaxed with two different methods: external potential and WVT, for three different resolutions as listed above.
Oscillations represent a superposition of a few lowest global oscillation modes that are slowly being damped by artificial viscosity.
Note that the WVT-relaxed configuration has a much higher oscillation amplitude because of the initially worse deviation from the target density profile.
As expected, higher particle numbers lead to smaller oscillation amplitudes and less damping.
The former is because more particles allow for more resolution to represent the density.
The latter is because viscosity acts as a local damping mechanism operating within the particle smoothing length (Eq.~\ref{eq:mg_visc}) that decreases with resolution.

\begin{figure*}[htp!]
\begin{tabular}{cc}
  \includegraphics[width=0.485\textwidth]{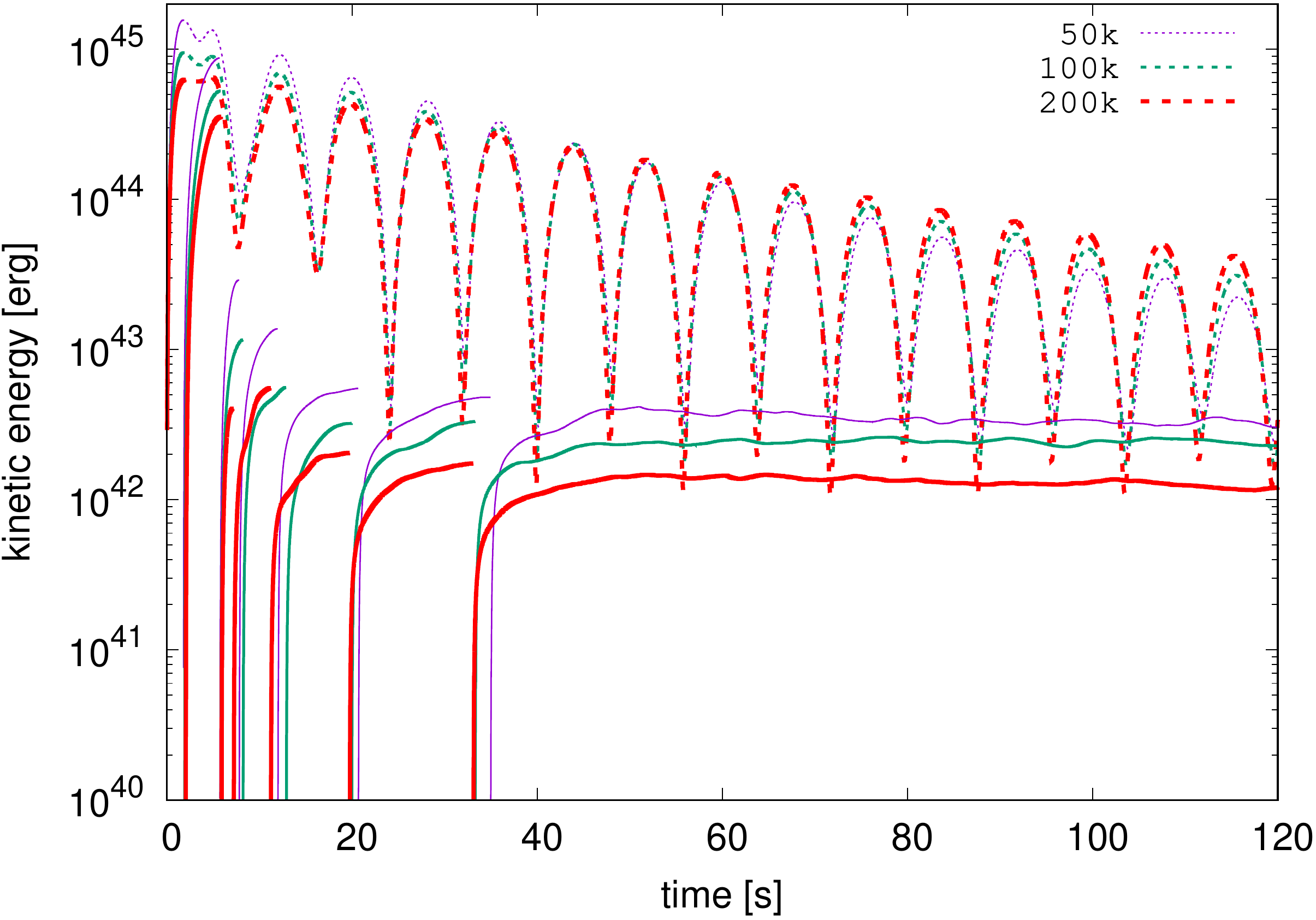} &
  \includegraphics[width=0.485\textwidth]{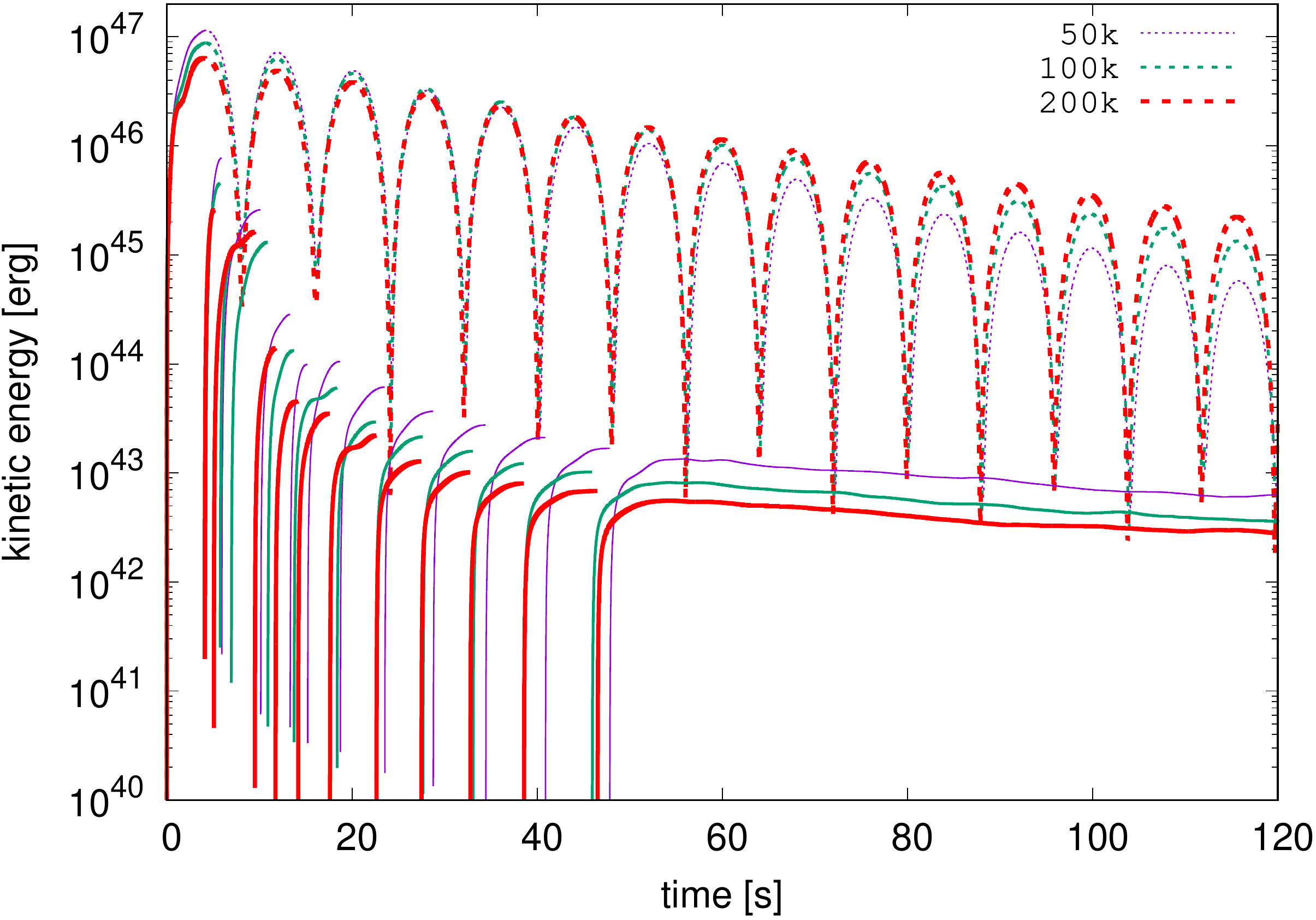}
\end{tabular}
\caption{Left: kinetic energy evolution of a particle configuration, relaxed with the effective potential, with HPR (solid lines) and without HPR (dashed lines), for three different resolutions.
Right: the same for particle configurations, relaxed with WVT method.
}
\label{fig:hprevo}
\end{figure*}

\begin{figure*}[htp!]
\begin{tabular}{cc}
  \includegraphics[width=0.485\textwidth]{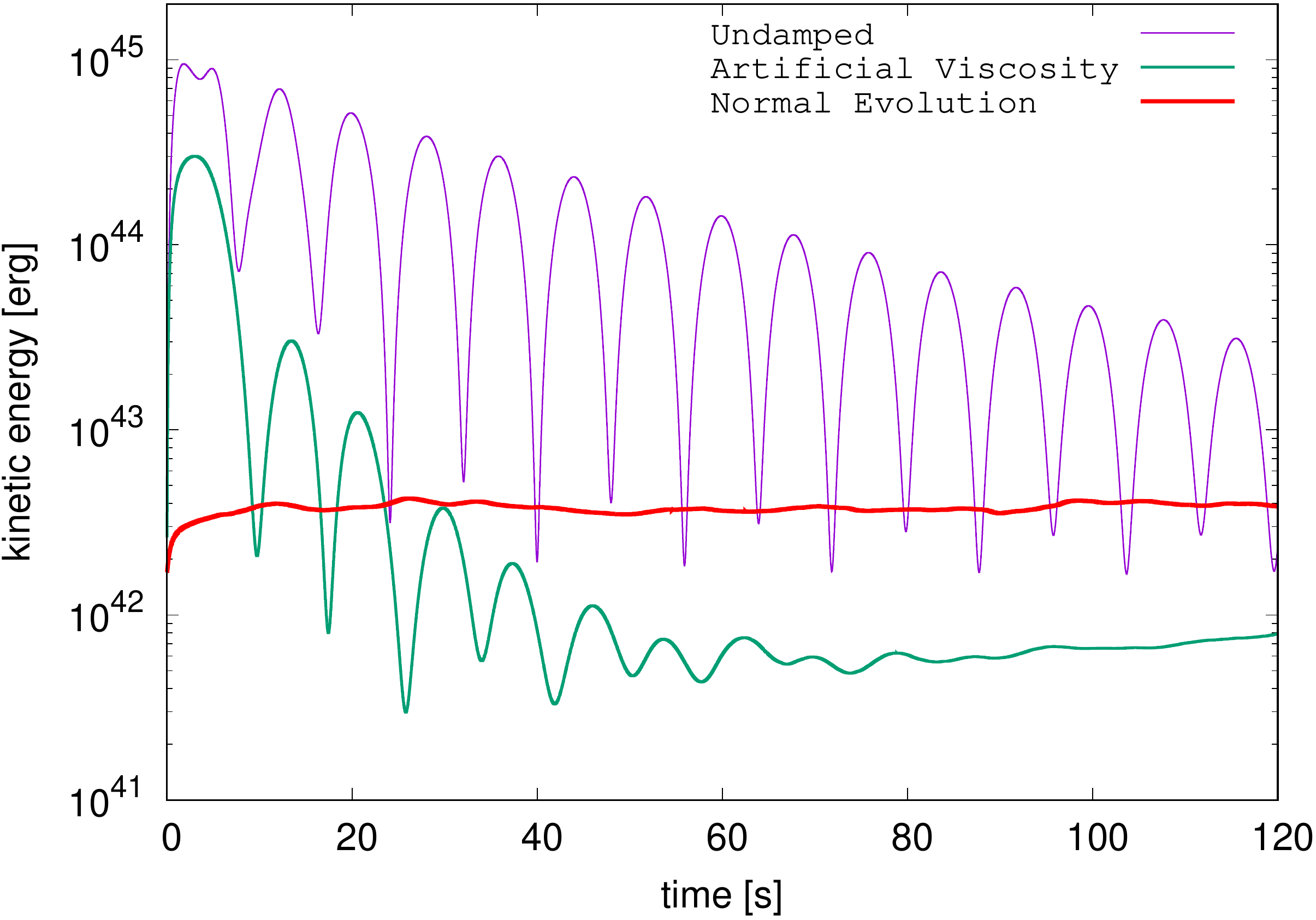} &
  \includegraphics[width=0.485\textwidth]{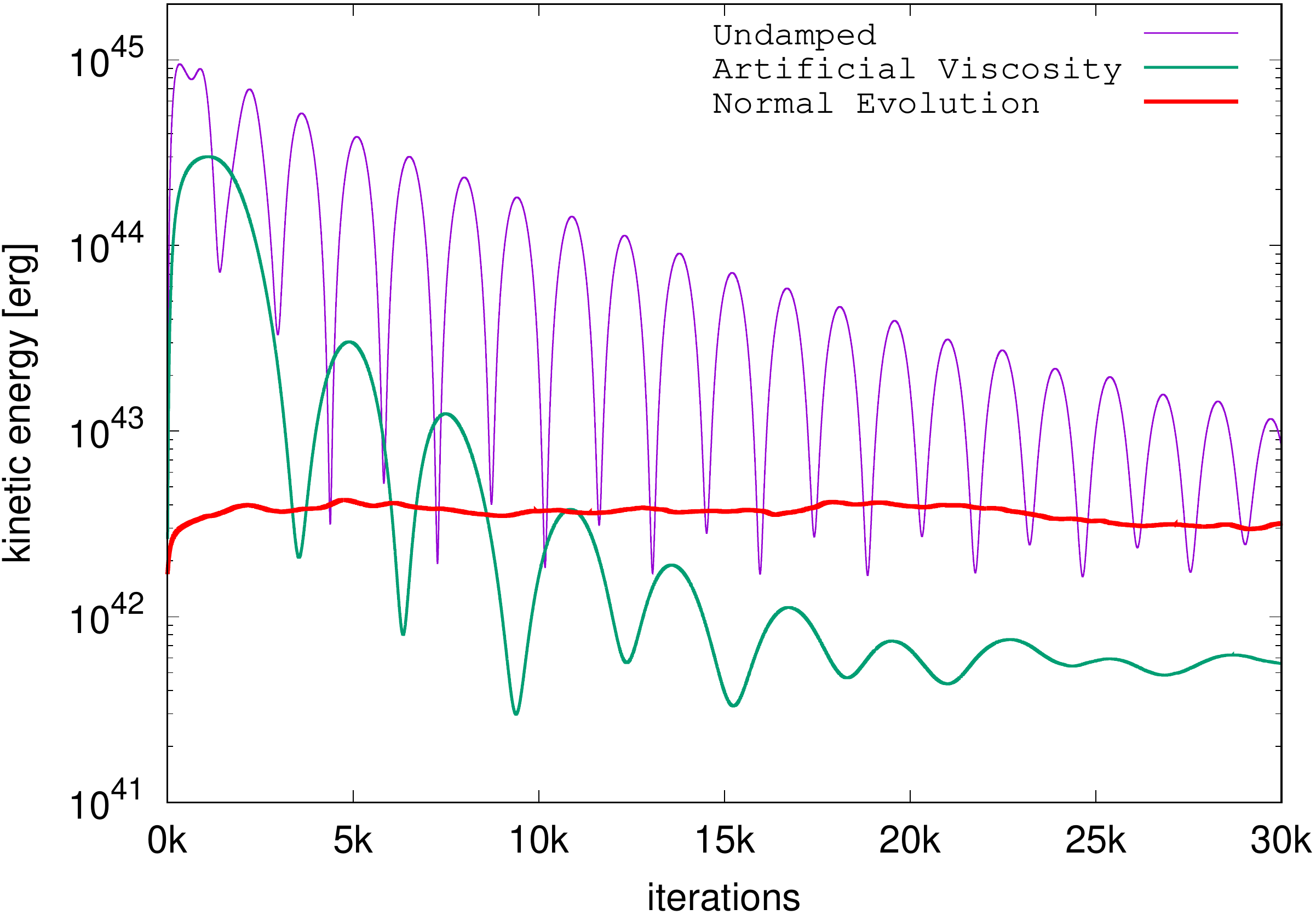}
\end{tabular}
\caption{Artificial viscosity relaxation for the 100k-particle configuration, relaxed in the external potential.
The normal evolution curve is recovered after returning the viscosity to a physical value and running without damping. The number of iterations required to recover a kinetic energy floor of the same order as the HPR damped stars is roughly double.}
\label{fig:artvisc}
\end{figure*}

We apply two methods to remove the global modes from the relaxed stars: HPR and artificial viscosity. 
The results are shown in Figures~\ref{fig:hprevo}~and~\ref{fig:artvisc}.
Both the HPR and artificial viscosity methods remove the global breathing mode.
However, the artificial viscosity method needs about twice as many iterations to reach a stable configuration at 120 seconds compared to HPR ($4.3 \times 10^4$ versus $2.2 \times 10^4$ iterations for the $10^5$ particle star). 
After viscosity is returned to typical values, kinetic energy rebounds to similar levels as obtained with HPR.
Note that the efficiency of the artificial viscosity method to suppress global oscillations is expected to be diminished with a higher number of particles.
This is similar to the resolution trends observed in Figure~\ref{fig:global-breathing-mode}, where oscillations of higher-resolution models decay slower.

\begin{figure}
  \includegraphics[width=0.48\textwidth]{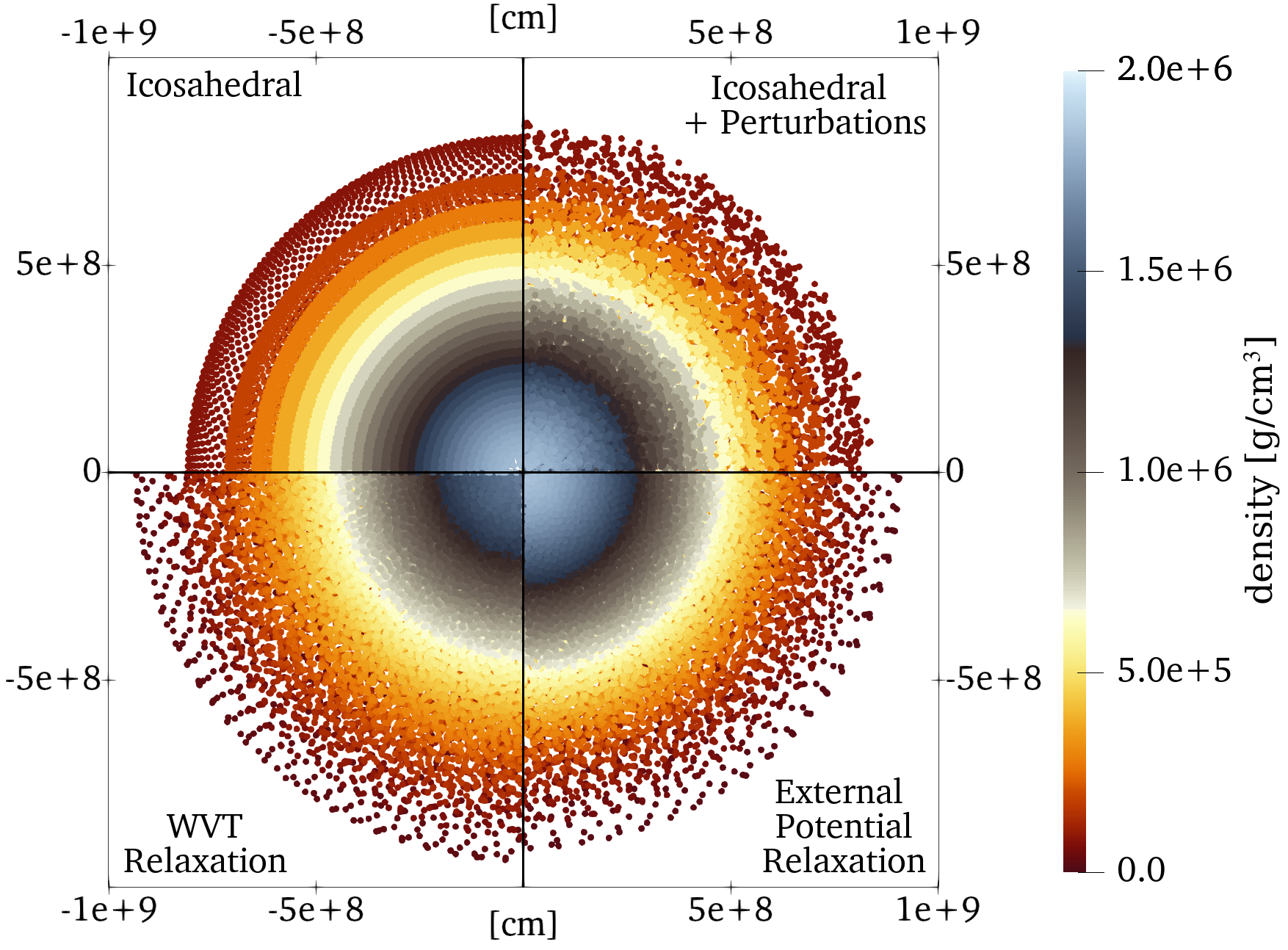} 
\caption{Comparison of the density cross-section of a $0.5\:M_{\odot}$ WD with $10^5$ particles for the purely icosahedral initialization, the icosahedral distribution with position perturbations, the particle distribution after external potential, and WVT relaxations.}
\label{fig:particle-configurations}
\end{figure}

Figure~\ref{fig:particle-configurations} provides a visualization of the density distribution for the initialization and relaxation of WDs with $\:0.5\:M_\odot$ and $10^5$ particles. 
It can be seen that the WVT-relaxed star is an outlier with a larger radius and lower central density in comparison to stars produced with the other relaxation methods. 
Again, this should not be viewed as a critique of the method but rather a note that our implementation in FleCSPH most likely requires improvement. 
\begin{figure}
  \includegraphics[width=0.48\textwidth]{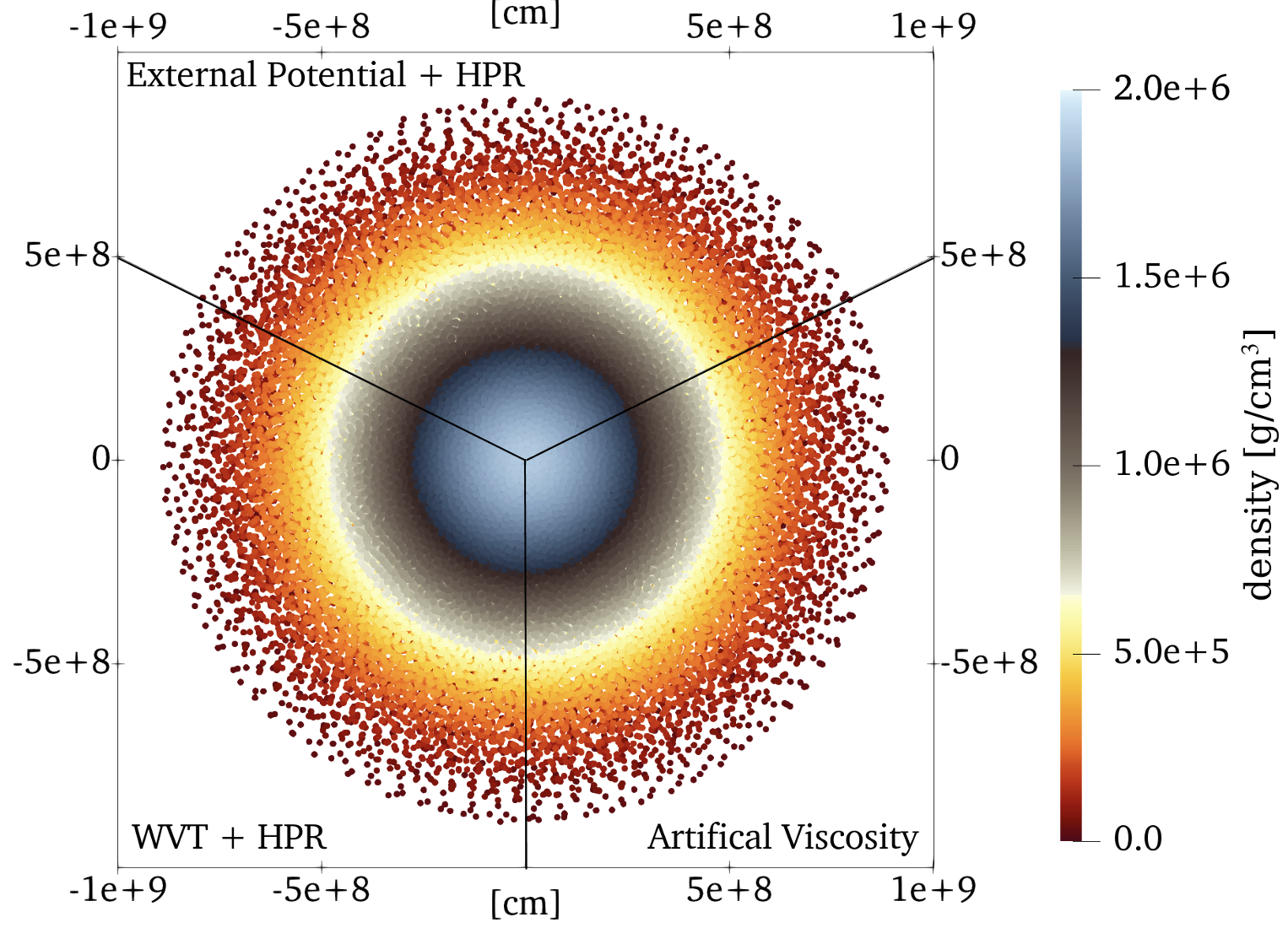} 
\caption{Comparison of the density cross-section in a $0.5\:M_{\odot}$ WD with $10^5$ particles after HPR and artificial viscosity damping of global modes.}
\label{fig:particle-hpr}
\end{figure}

\subsection{White Dwarf-White Dwarf Binaries}
WDB systems survive for up to billions of years, during which they slowly lose their angular momentum to gravitational waves (GW) until the eventual merger. 
Macro characteristics of the binary, e.g., component masses and mass ratio, determine the final stages of this process and whether the merger will be energetic enough to produce a Type Ia supernova. 
\edt{Here, we briefly describe the conditions a binary must satisfy to trigger unstable mass transfer and a complete merger. These will also be the initial conditions for our simulation setup.}{For the work which follows, we wish to set up and study WD binary systems that will undergo unstable mass transfer, resulting in complete tidal disruption of the donor WD. 
The process of mass transfer becoming unstable is closely tied to initial conditions.
Here we choose the binary stars' masses taking into account two distinct arguments.
The first one is the analytic model for which binary mass characteristics should be unstable, and the second argument is historical: the adopted mass ratio has been studied in the past.~\citep[e.g.][]{DSouza2006,Motl2017}}

The emission of GWs effectively describes the evolution of WDB orbits. 
Once GWs have driven the binary to a critical separation called the Roche-Lobe Overflow (RLOF) separation or $a_{\mathrm{RLOF}}$, mass from the donor, which has over-filled the Roche potential, begins to transfer to the accreting WD. 
A well-known fit for the RLOF radius as a function of the binary mass ratio between stars 1 and 2 is given by:
\begin{eqnarray}
    \label{eq:mt1}
    R_{\mathrm{RL}} = a \: r_{\mathrm{RL}}(q) = a \: \frac{0.49\:q^{2/3}}{0.6\: q^{2/3}+\ln{(1+q^{1/3})}}\text{,}
\end{eqnarray}
where $q=m_{\mathrm{1}}/m_{\mathrm{2}}$ and $a$ is the orbital separation \citep{Eggleton1983}. 
Note that $a = a_{\mathrm{RLOF}}$ when $R_{\mathrm{RL}} = R_{\mathrm{WD}}$.

There are two scenarios for mass transfer.
The mass transfer is considered stable if it occurs on the same timescale as the GW emission. 
It requires the binary orbit to grow such that the donor star remains at the RLOF separation. 
The stability of mass transfer is maintained as long as the Roche Lobe remains filled.
With that, the timescale for the growth of the donor star's Roche-Lobe radius is equal to the change in the radius of the donor star, i.e., $\dot{R}_{\mathrm{wd}} = \dot{R}_{\mathrm{RL}}$.

If the donor star's radius grows much faster than the Roche-Lobe radius, that is $\dot{R}_{\mathrm{wd}} > \dot{R}_{\mathrm{RL}}$, then the mass transfer quickly becomes unstable and the donor WD will be tidally disrupted. 
This is possible for WDs since they are degenerate objects, and their radius grows as they lose mass. 
The timescale for such a merger is on the order of the orbital timescale of the pre-merger binary system. 

To find the critical-mass ratio for stability, one needs to determine the value of $q$ for which the following inequality is true: 

\begin{eqnarray}
\label{eq:unstablecriteria}
     \frac{d \: \ln \left(R_{\mathrm{wd}}\right)}{d \: \ln \left( m_{\mathrm{wd}}\right)}\:\frac{\dot{m}_{\mathrm{wd}}}{m_{\mathrm{d}}}&>&\frac{\dot{a}}{a} +\frac{d \: \ln \left(r_{\mathrm{RL}}\right)}{d \: \ln \left( q\right)} \: \frac{\dot{q}}{q}\text{.}
\end{eqnarray}
\edt{}{The response of the donor WD is included in the left-hand side of the above equation and is strongly dependent on the chosen equation of state.
For simplicity, the following analytics use a Zero-Temperature White Dwarf EOS \citep{Chandrasekhar1957} for determining the adiabatic response of the donor WD to mass loss.
The response of the accretor WD to mass accretion is ignored as the accretor's radius is much smaller than it's Roche radius.}

The change in the orbital separation with time is determined by conserving the change in the total angular momentum of the orbit. 
We take into account mass loss from the orbit, gravitational wave emission, and mass transfer between the two components with the following form:

\begin{align}
\label{eq:stablecriteria}
    \frac{\dot{a}}{a} = &\frac{\dot{\mathcal{J}}_{\mathrm{gw}}}{\mathcal{J}_{\mathrm{orb}}}\frac{2 \mathcal{J}_{\mathrm{orb}}}{\left(\mathcal{J}_{\mathrm{orb}} +\mathcal{J}_{\mathrm{d}}\right)}
     - \frac{\dot{m}_{\mathrm{wd}}}{m_{\mathrm{wd}}\left(\mathcal{J}_{\mathrm{orb}} +\mathcal{J}_{\mathrm{d}}\right)}\times\nonumber\\
    \times&\left[2 \mathcal{J}_{\mathrm{orb}} -2(1-\beta)\left(q \mathcal{J}_{\mathrm{orb}} + \frac{m_{\mathrm{wd}}}{m_{\mathrm{d}}}\mathcal{J}_{\mathrm{d}}\right)\right. \nonumber\\
    &+\left.\frac{d\ln r_{\mathrm{d}}}{d \ln q}\mathcal{J}_{\mathrm{d}}-\beta \frac{q}{q+1}\mathcal{J}_{\mathrm{orb}}\right]\text{.}
\end{align}

$\mathcal{J}$ is the angular momentum for each component labeled with the subscript: the subscript $\mathrm{d}$ denotes the disk, and $\mathrm{orb}$ denotes the orbit. 
$\beta$ is a parameterization for a fraction of material that is ejected from the local system. 
$r_{\mathrm{d}}$ is the radial fraction of the disk assumed to be $R_{\mathrm{d}} = r_{\mathrm{d}}\: a$. 

\begin{figure}
  \includegraphics[width=0.485\textwidth]{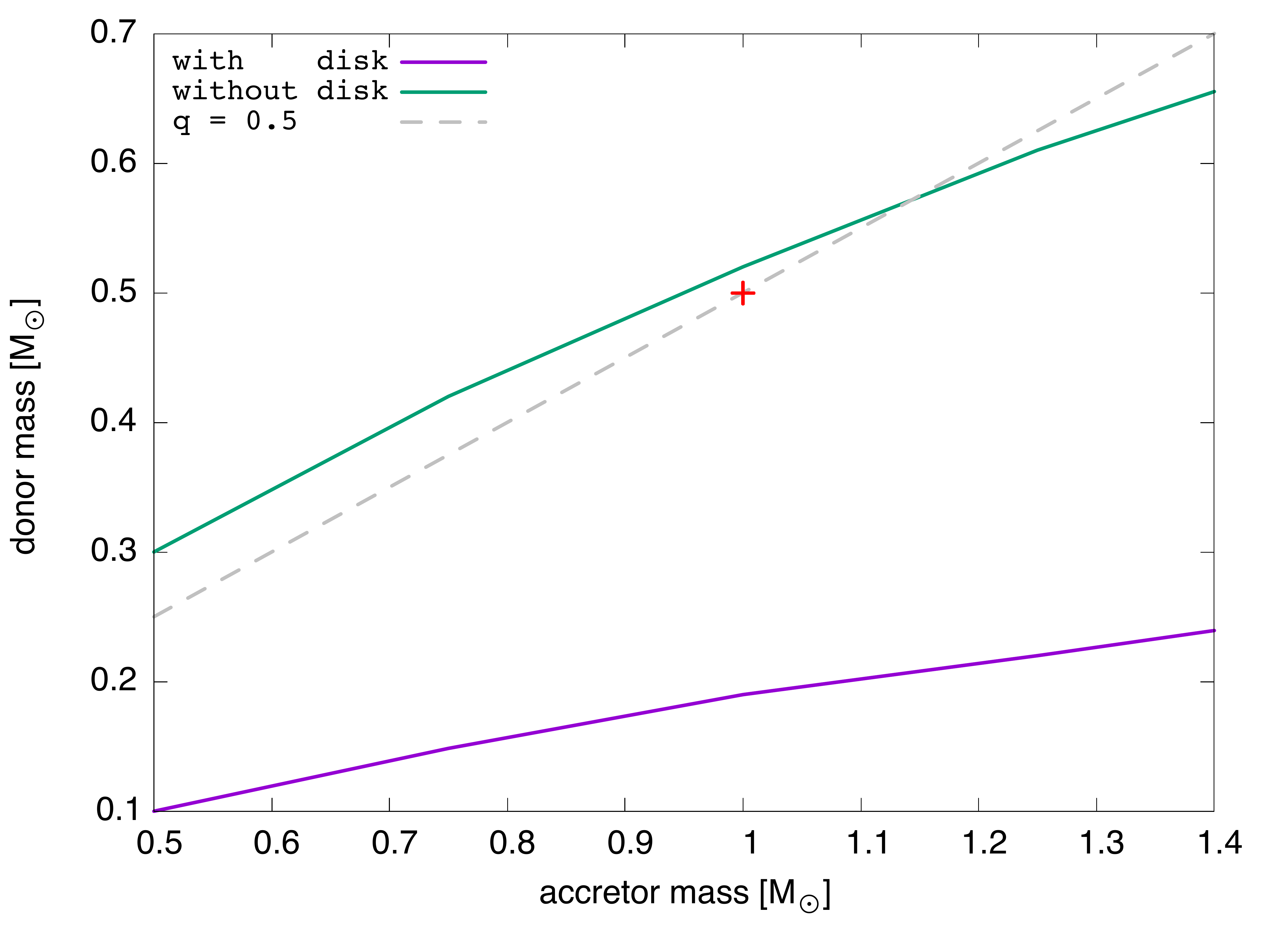} 
\caption{\edt{}{The critical instability lines for a binary system with and without the formation of an accretion disk (systems above the critical values for the donor mass are unstable). 
We see that when the binary undergoes direct accretion with no mass loss, the unstable critical $q$-value is much higher. The grey dashed line represents the location of $q=0.50$, and the red point indicates our chosen binary.}}
\label{fig:linestability}
\end{figure}

\edt{We find the line of maximum stability}{The line of maximum stability is determined} by solving Eq.~\ref{eq:unstablecriteria} and determining when the left-hand side is equal to the right-hand side for different masses of the accreting star.
When the accretor has a sufficiently large radius, e.g., in the case of a BWD, the accreting material is not deflected enough to form a disk\edt{; instead, the material directly transfers onto the accretor}{}.
\edt{We find that mass transfer becomes unstable for $q \geq 0.4$.
For a more detailed analysis, we refer the reader to [Fryer1999a], who discuss the intricate stability line for similar systems.}{The critical instability lines are shown in Figure~\ref{fig:linestability}.}

\edt{}{
\cite{Motl2017} studied a significant number of binaries in both grid-based and SPH simulations.
They found that the binaries with $q=0.5$ may or may not merge, depending on other parameters of the system.
\cite{DSouza2006} showed that the long stability of the binary with $q=0.5$ is strongly tied to the driving mechanism.
Our analytic model (\ref{eq:unstablecriteria},\ref{eq:stablecriteria}) suggests that when accretion is direct, this mass ratio exists right at the edge of stability.
This uncertainty makes the $q=0.5$ binary optimal for our purpose.}

\edt{For our double WDs setup}{Thus}, we chose a binary with $q=0.5$. 
\edt{For the single relaxed stars, we additionally relax a companion}{We pick an accretor}
with $1 \: M_\odot$ and $10^5$ particles \edt{}{and a donor WD with varying resolutions}, producing three binaries of $1.5 \times 10^5$, $2 \times 10^5$, and $3 \times 10^5$ particles. 
Finally, we also prepare a binary with $3000$ particles per star following the methods above and use these WDs in binary systems for testing.
\subsubsection{Roche Relaxation}
\label{sec:3.2.1}
\begin{figure}[htp!]
\centering
  \includegraphics[width=\columnwidth]{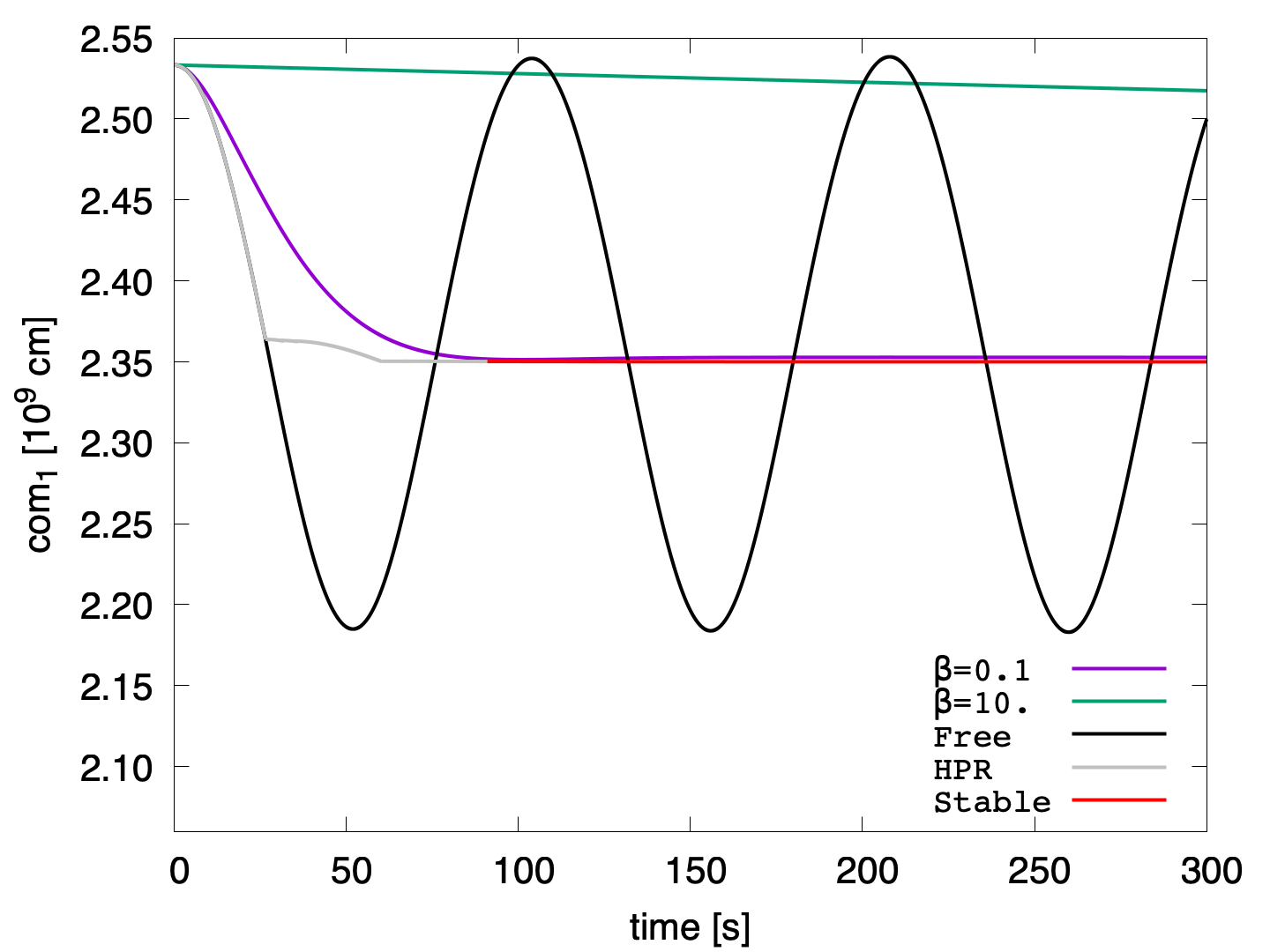}
\caption{Position of the center of mass of one of the stars on the $x$ axis as an indicator of binary relaxation, during the first 300 seconds of particle relaxation run, for the test 6k-particle binary.
The red line indicates the target value of the stable center-of-mass coordinate.
Here we see the effectiveness of HPR in quickly driving the binary to the stable configuration. HPR is compared with the same system when it is allowed to evolve (without velocity damping) freely and with damping via a small drag force ($\beta=0.1\ {\rm s}^{-1}$) and a large drag force ($\beta=10\ {\rm s}^{-1}$). The time to suppress spurious oscillations for $\beta=0.1\ {\rm s}^{-1}$ looks similar to the time for the HPR damping; however, the HPR reached stability at 4500 iterations, while the velocity damping method required 7500 iterations to approach the exact location of the center of mass.}
\label{fig:damping}
\end{figure}

\begin{figure*}[htp!]
\begin{tabular}{cc}
  \includegraphics[width=0.485\textwidth]{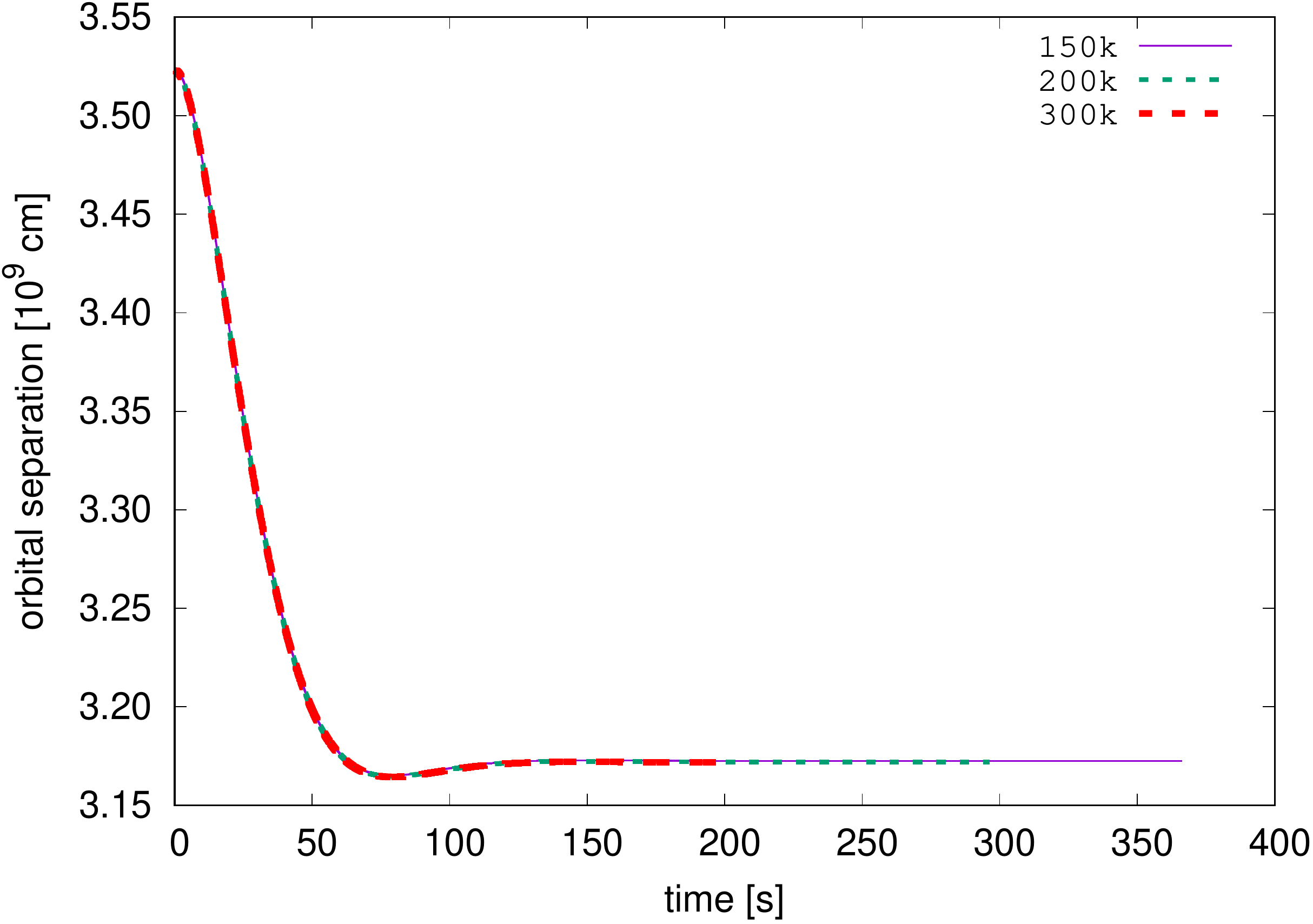} &
  \includegraphics[width=0.485\textwidth]{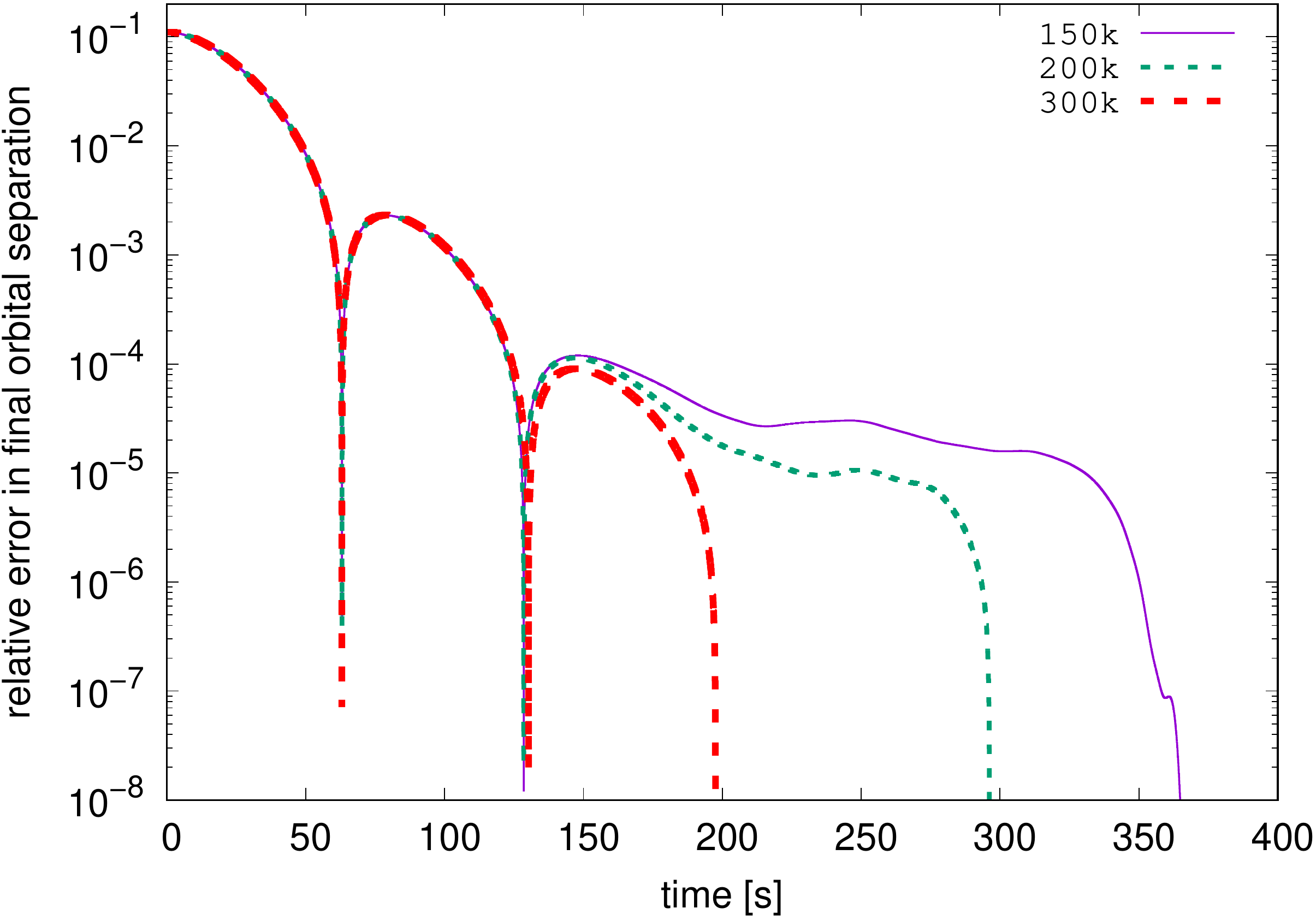}
\end{tabular}
\caption{Left: evolution of the orbital separation of a WDB system at three particle resolutions, as the system is driven towards the target separation, 105\% $a_\mathrm{RLOF}$, using the relaxation method described in Section \ref{sec:3.2.1}.
Right: relative error with respect to the final separation in the log plot.
After all orbital oscillations have been effectively removed from a wide system, we find that the binary separation converges to the target value with relative accuracy of $10^{-8}$.}
\label{fig:relaxing-to-separation}
\end{figure*}

We produce our desired binary systems by combining individually relaxed stars at a given orbital separation $a_\mathrm{init}$. 
The system is then evolved in the co-rotating frame with an effective potential as described in Section~\ref{sec:extpotrel}.
However, typically when we combine two individually relaxed stars into a binary, new perturbations are introduced due to tidal forces and self-gravity.
To address these perturbations, we apply HPR individually to each star in the binary, monitoring individual kinetic energies and freezing the velocities when either one reaches maximum.
Because the stars are not point masses, the binary still possesses a significant spurious initial eccentricity, creating oscillations in the orbital separation with an amplitude of around 10\% $a_\mathrm{init}$. 
We also found HPR to be very effective at damping out this eccentricity. 
Thus, once individual breathing modes have been sufficiently removed, we also apply HPR to the total kinetic energy of the system (see Figure~\ref{fig:damping}). 
With the well-relaxed initial system, we drive the binary by the target angular momentum toward the desired separation. 
During this stage, we use a particle drag force that works against the particles' velocities and is computed as $\vec{a}_{\mathrm{drag}} = -\beta \vec{v}$ (see also Section~\ref{sec:extpotrel}). 
Once the system has reached the desired separation, we find it free of oscillations, as demonstrated in Figure~\ref{fig:relaxing-to-separation}. 

\begin{figure}[htp!]
 \centering
 \includegraphics[width=\columnwidth]{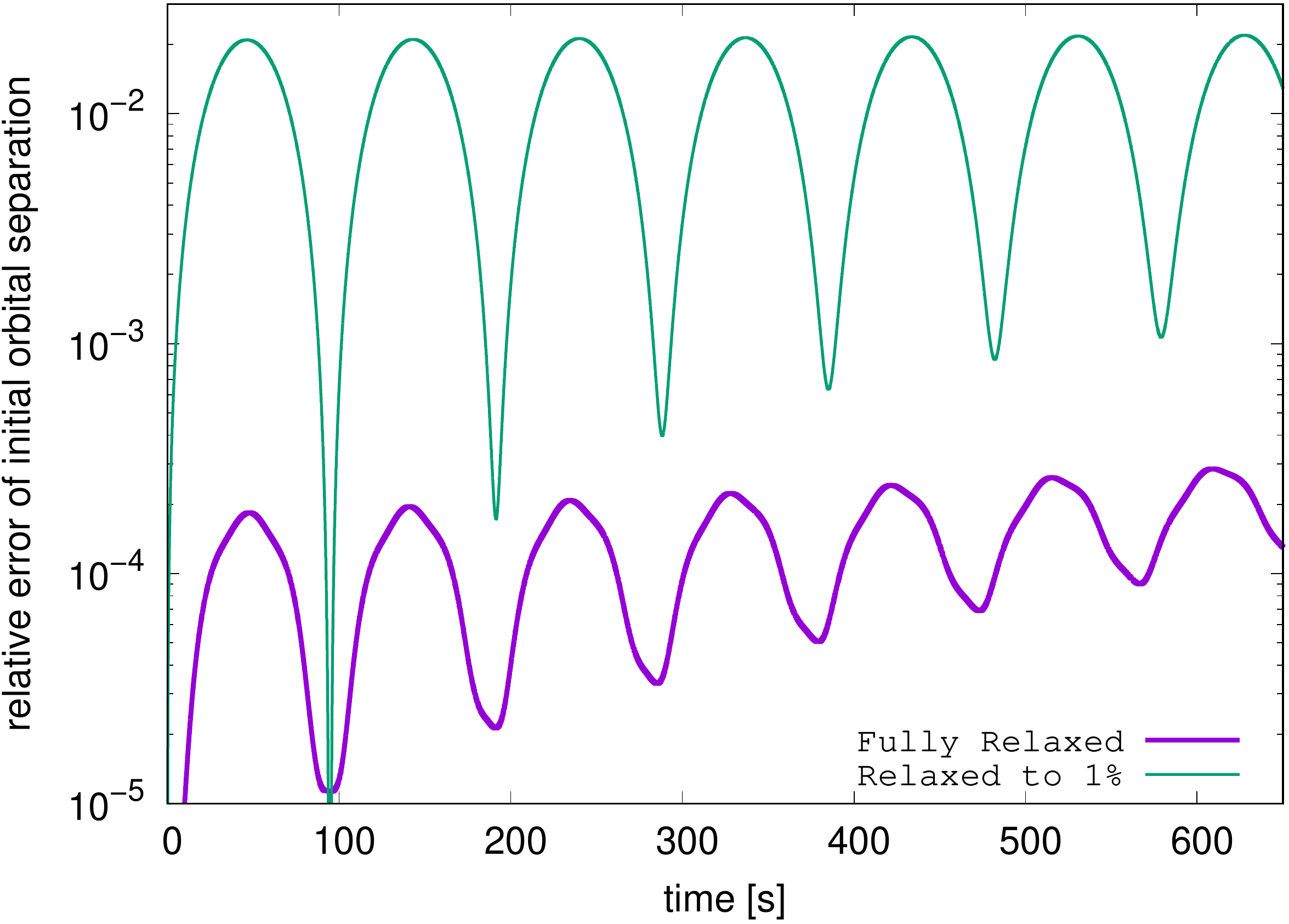}
\caption{Orbital separation as a function of time for two 6k-particle WDB configurations: fully relaxed (thick purple line) and relaxed to 1\% of the converged separation (thin green line).
Here, we see that a 1\% error in orbital separation results in a 2\% oscillation mode, while for the fully relaxed system, oscillations only reach a much smaller value of $\sim0.02\%$.}
\label{fig:relaxed-vs-not}
\end{figure}

After the binary has converged to a stable separation, we proceed to map the orbital velocities to the particles and run the full hydrodynamic simulation.
A minor oscillation emerges in the orbital separation with the same frequency as the orbit of the binary, as shown in Figure~\ref{fig:relaxed-vs-not}, manifesting a spurious eccentricity. 
Since such oscillations were not seen in the co-rotating frame, we assume that they are numerical in nature and an artifact of the FMM scheme for approximating Newtonian gravity. 
The latter uses a so-called ``minimal acceptance criteria'' (MAC) angle to group distant particles for bulk interaction \citep[see, e.g.,][]{Dehnen2014}. 
As the particles move in their orbit, the FMM nodes are computed with slight variation, which imposes the small orbital oscillation as the system changes its angular orientation.

\textbf{Wide-Orbit Stability:}
To test the relaxation method and conservation in our binaries, we began by choosing an orbit for the 6k-particle system that should remain stable in our SPH simulations at initial separation $a = 1.25 \: a_\mathrm{RLOF}$. 
In order to compare the effectiveness and relevance of accurate relaxation, we relaxed the 6k-particle system to be within one percent of the converged value. 
Figure~\ref{fig:relaxed-vs-not} compares a fully and a nearly relaxed system. 
We found that the fully relaxed system exhibits a small residual eccentricity, presumably due to the FMM approximation, on the order of $10^{-4}$ of the orbital separation, which amounts to a 7~km deviation for a circle. 
\begin{figure*}[htp!]
\begin{tabular}{cc}
  \includegraphics[width=0.50\textwidth]{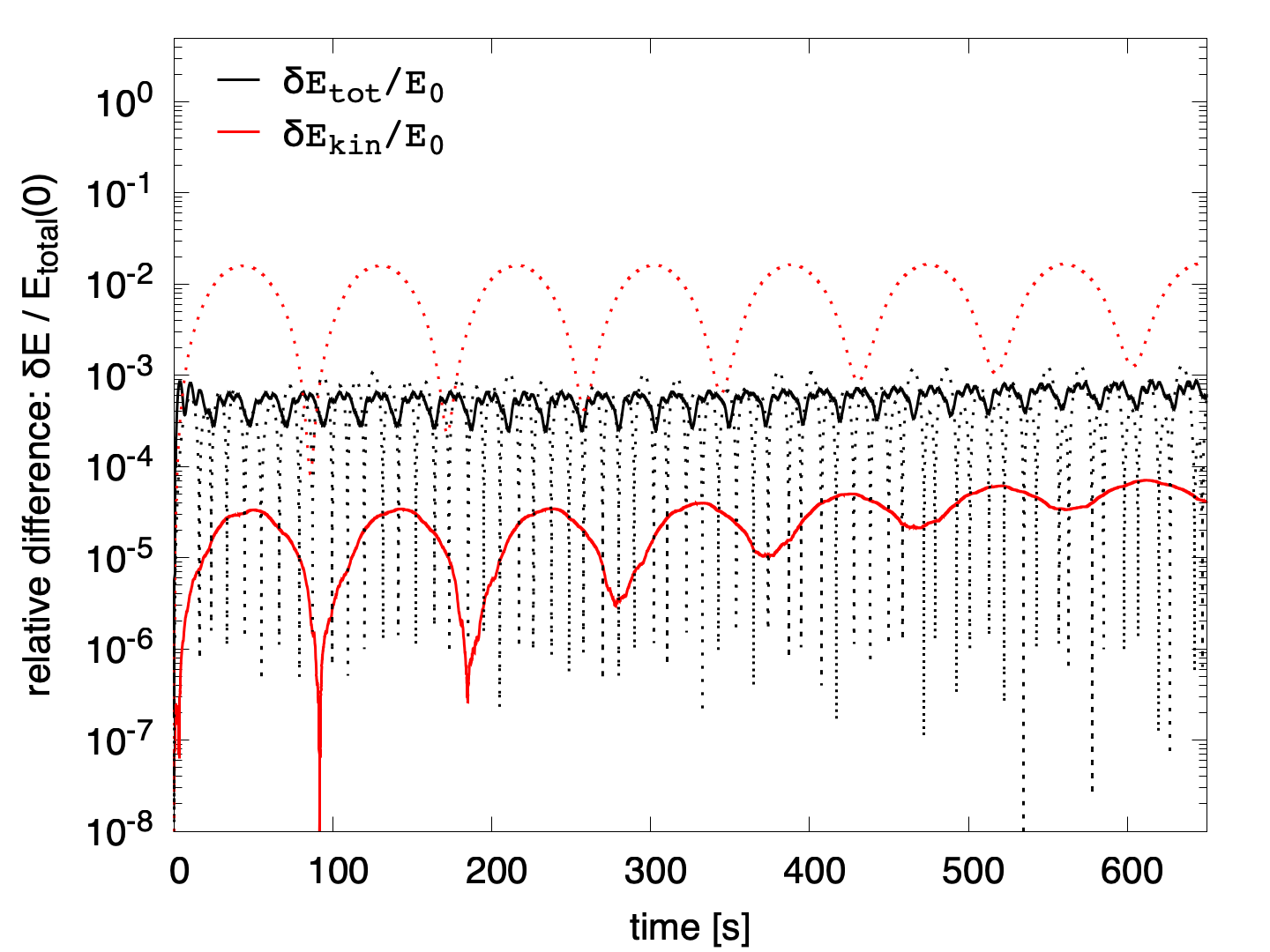} &
  \includegraphics[width=0.50\textwidth]{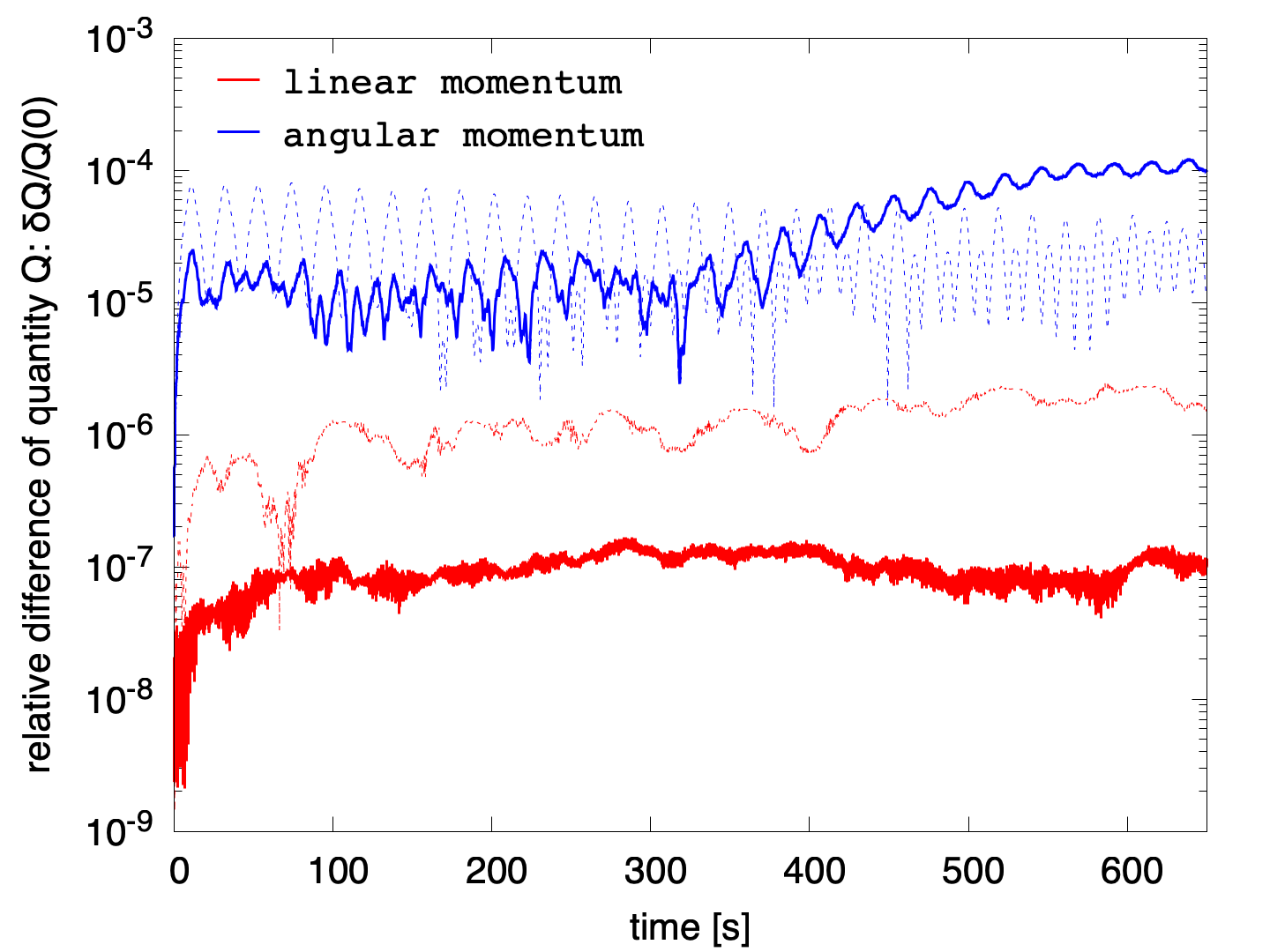}
\end{tabular}
\caption{
Evolution of the relative difference of the conserved quantities.
Left: evolution of the difference in energies, total and kinetic, for the 6k particle wide binary: nearly relaxed (dashed line) and fully relaxed (solid line) binary. 
The differences are divided by the initial value of total energy for each of the systems.
Note that the total energy is less than the kinetic because it includes negative gravitational potential energy.
Right: relative difference in the linear (red) and angular (blue) momenta for the nearly relaxed (dashed line) and fully relaxed (solid line) binary.
The difference is divided by the initial value of the corresponding quantity.}
\label{fig:conservation}
\end{figure*}
The system relaxed only to 1\% shows a residual orbital eccentricity on the order of $10^{-2}$, or 700~km deviation from the circle.
Figure~\ref{fig:conservation} demonstrates the conservation properties of the binary over five orbits.

\textbf{Roche Lobe Overflow:}
When the desired orbital separation is at the critical value, we must more diligently damp out any oscillatory modes, as these can have a dramatic effect causing unphysical mass transfer before the critical separation is found or a possible stopping of the accretion stream. 
Either of these effects will completely distort the accretion dynamics. 
In order to find the critical separation, we relax the systems to 105\% $a_\mathrm{RLOF}$ following the method above. 
When we have converged to this separation, we increase the drag force on the system as we target the $a_\mathrm{RLOF}$ separation. 
\begin{figure*}[htp!]
\centering
  \includegraphics[width=0.95\textwidth]{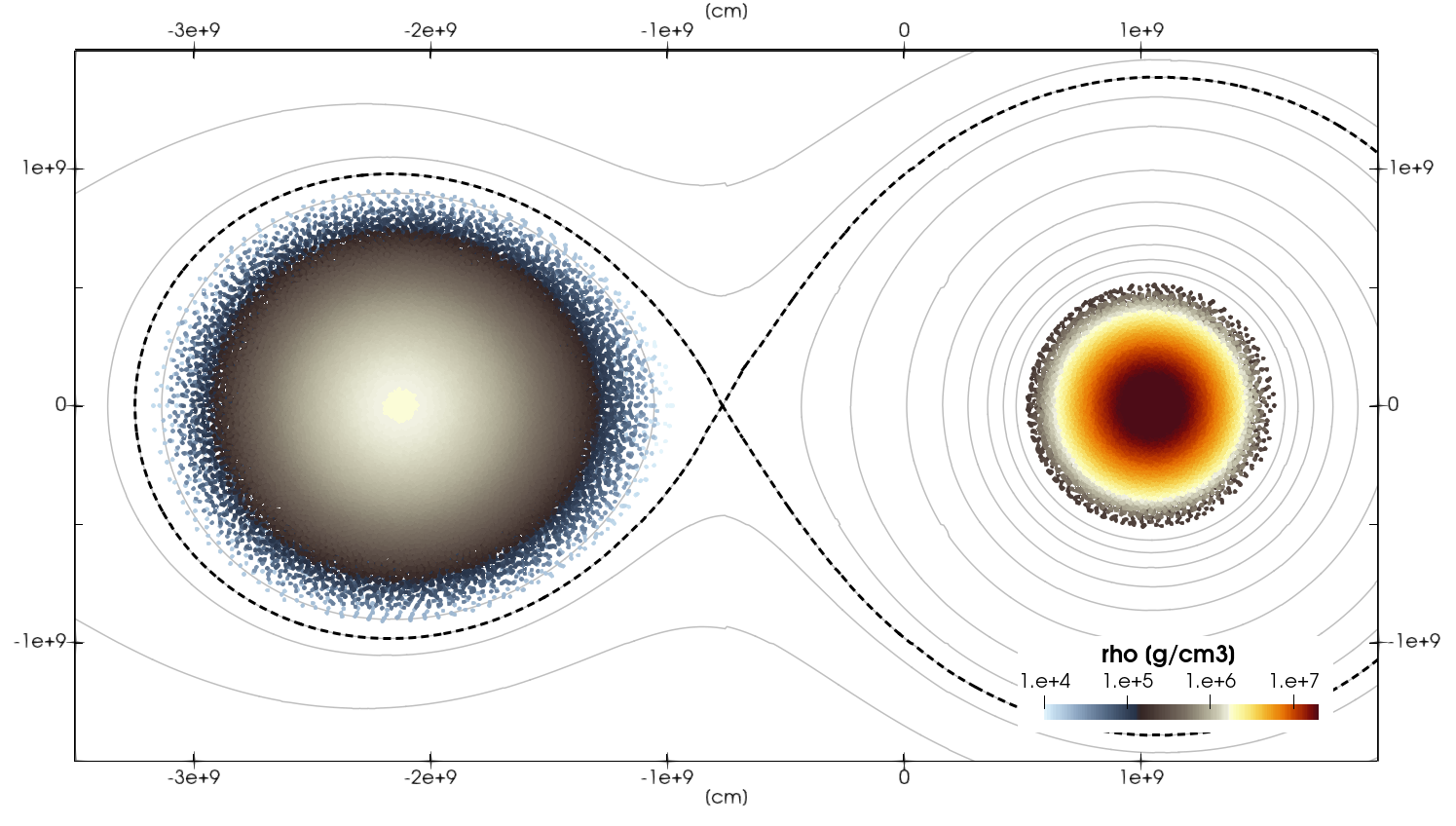}
\caption{Snapshot for the 300k resolution binary system relaxed to just before their critical separation distance. The contours of the effective potential are, counting from the outermost contour, at $(-1.1, -1.2, -1.243, -1.3, -1.4, -1.6, -1.8, -2.0, -2.2, -2.4, -2.6)\times10^{17}\ {\rm cm}^2\,{\rm s}^{-2}$. The dashed line marks the location of the Roche lobe in the potential.}
\label{fig:10}
\end{figure*}
Once the first particle overflows through the L1 point, we stop the relaxation and map the orbital velocities at this orbital separation to the particles. 
Figure~\ref{fig:10} shows the binary at $a_\mathrm{RLOF}$ for the $3\times 10^5$ system.

\edt{\textbf{Accretion Streams:}
All the relaxation schemes aim to simulate the merger of compact binaries while accurately modeling the initial accretion stream. To that end, we show the resulting accretion streams from the three resolutions of binaries after full relaxation and orbital velocity mapping, see Figure [fig:11]. As the particle number increases for the 0.50 $M_\odot$ WDs in the binary, the resolution of the stellar atmosphere benefits significantly. 
In particular, the region of the donor star which feeds the initial accretion stream is more accurately simulated. This underlines the importance of high-fidelity simulations and the need for highly efficient relaxation schemes.}{}

\subsection{Effect on dynamical evolution}
To accentuate the importance of accurate relaxation on the dynamics of a WDB merger, we present two setups of 150 thousand particles near the critical orbital separation, prepared with two different methods: with HPR relaxation and a generic method without HPR. 
In both cases, the HELMHOLTZ EoS is used for evolving the system \citep{Timmes2000}. 
The same EoS is used to prepare the initial configuration at an initial temperature of $10^5$~K\edt{ }{\textemdash{}under the assumption that the binary is considerably old and the components have had ample cooling time\textemdash{}}and a composition that consists of 49.5\% $^{12}$C, 49.5\% $^{16}$O, and 1\% $^4$He, which amounts to the average ion charge and atomic mass of $\bar{Z}=6.69$, $\bar{A}=13.38$ respectively. 

In the HPR setup, we use the previously described procedure to drive the binary to an orbital separation as close to the critical as we could achieve without particle transfer. 
We settle the binary in a circular orbit with an orbital separation of $3.1696\times10^9$~cm to a high degree of accuracy such that the evolution of the relative separation at the end of the relaxation phase does not exceed $\Delta a_{\rm orb}/a_{\rm orb}\simeq 10^{-8}$ (similar to what is shown in Figure~\ref{fig:relaxing-to-separation}).

In the generic setup, single isolated WDs are initialized and relaxed for a few dynamical times, $\tau_{\mathrm{dyn}}$ (usually on the order of $\sim10$~s for WDs).
We chose this relaxation time to be 30~s. 
In such a setup, global breathing modes are not explicitly suppressed and remain present to some extent.
After the individual relaxation, the two WDs are assembled in a wide binary to ensure that the tidal interaction is weak. 
The separation is typically set to be between 110\% and 180\% of the analytic prediction for $a_{\mathrm{RLOF}}$. 
We settled on the value of 150\%, i.e., we started from the wide-binary separation of $1.5 a_{\mathrm{RLOF}}$.

At this point, the binary must be artificially driven towards the Roche lobe overflow limit. 
While it is possible to drive the system in an inertial frame, we use a co-rotating frame for this procedure. 
The driving method varies in the literature \citep{Segretain1997,Guerrero2004,Motl2007,Yoon2007,Fryer2008,Raskin2012,Zhu2013,Dan2014,Raskin2014,Motl2017}, but what remains common is that angular momentum is drained from the binary, gradually decreasing the orbital separation.
This is usually done sufficiently slowly, over many dynamical times $\tau_{\mathrm{orb}}$ of the binary, which can be quite expensive computationally. 
While sufficiently gradual driving does suppress various oscillations, the ringing from low-frequency modes is only partially removed. 
We chose to drive the binary over three orbital periods.

\begin{figure*}[htp!]
\centering
\begin{tabular}{lr}
  \includegraphics[width=0.34\textwidth]{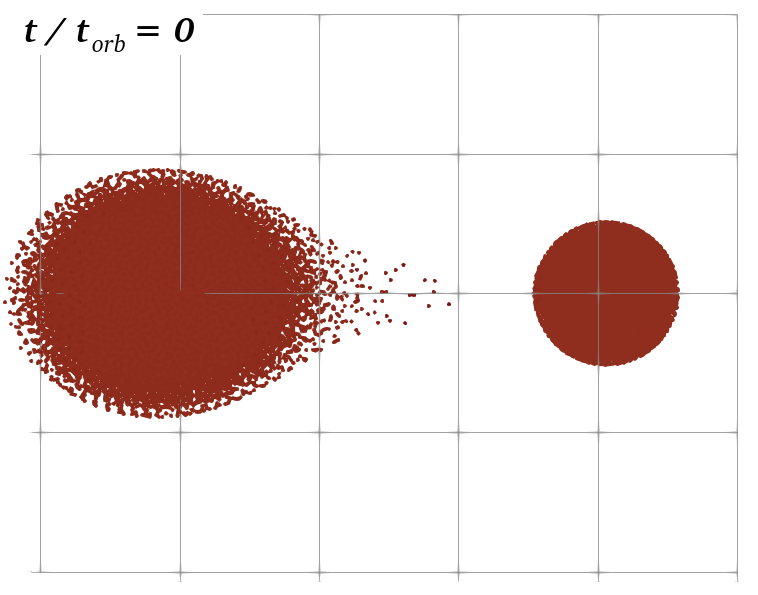} $\qquad\qquad$ &
  \includegraphics[width=0.34\textwidth]{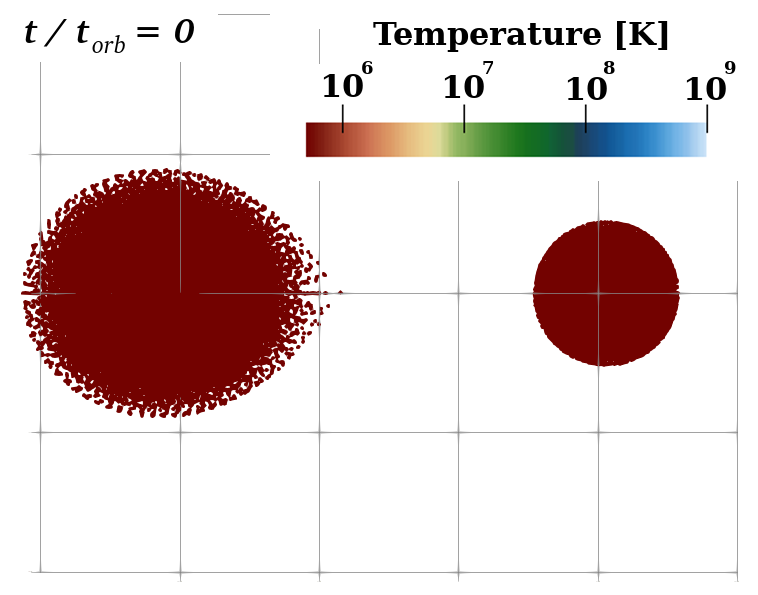} \\
  \includegraphics[width=0.34\textwidth]{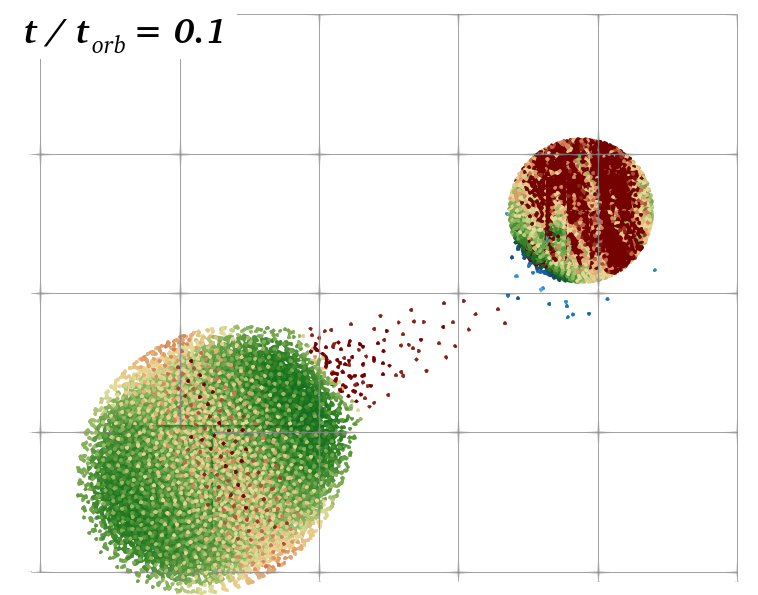} &
  \includegraphics[width=0.34\textwidth]{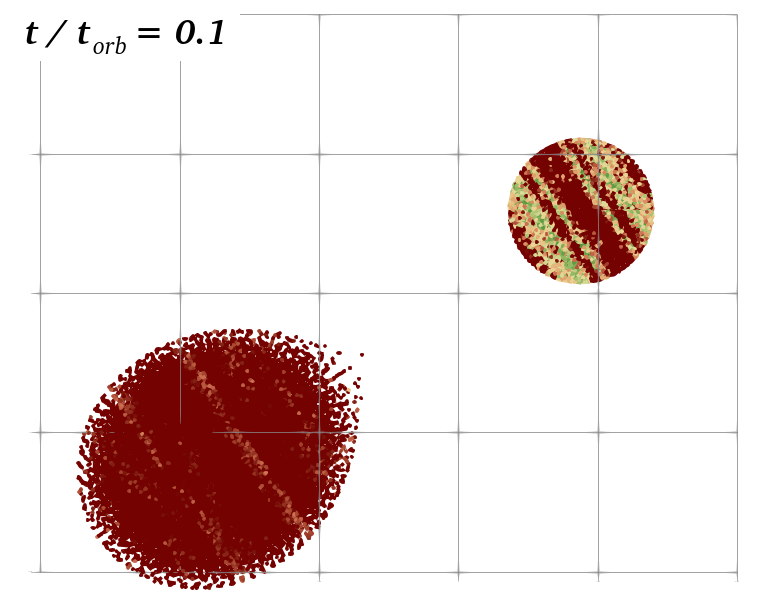} \\
  \includegraphics[width=0.34\textwidth]{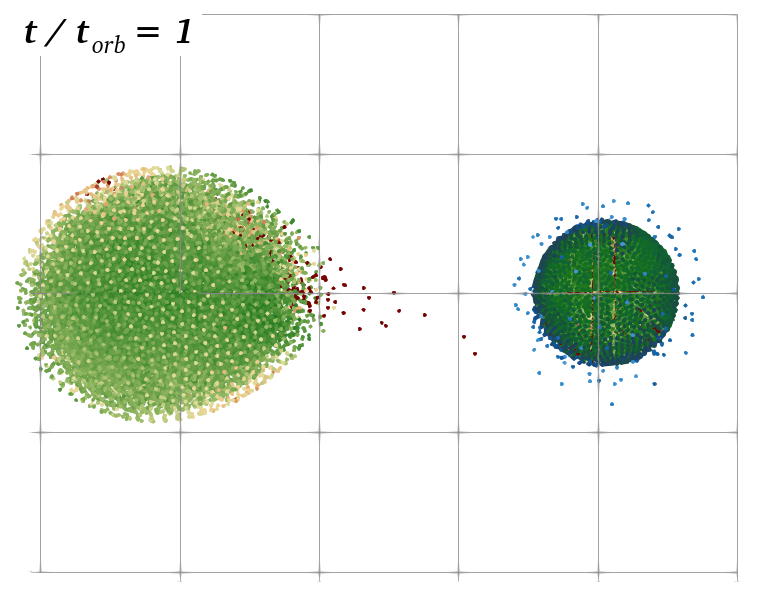} &
  \includegraphics[width=0.34\textwidth]{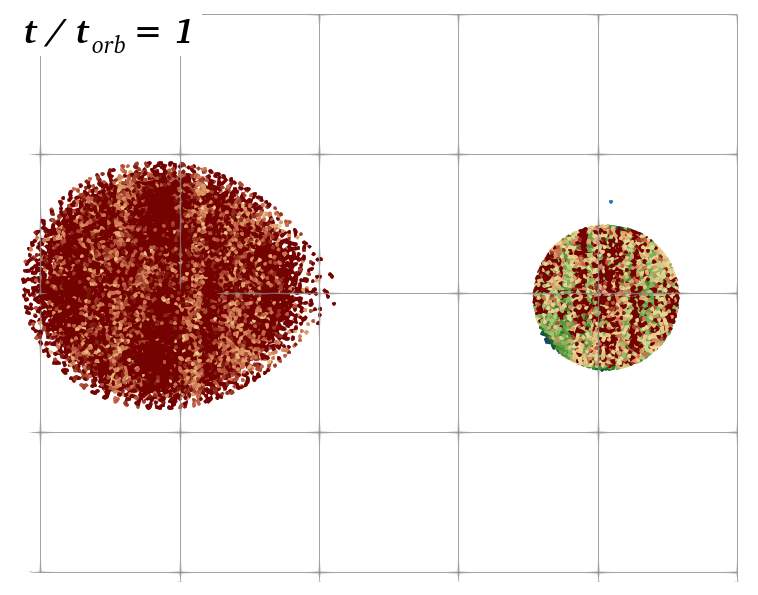} \\
  \includegraphics[width=0.34\textwidth]{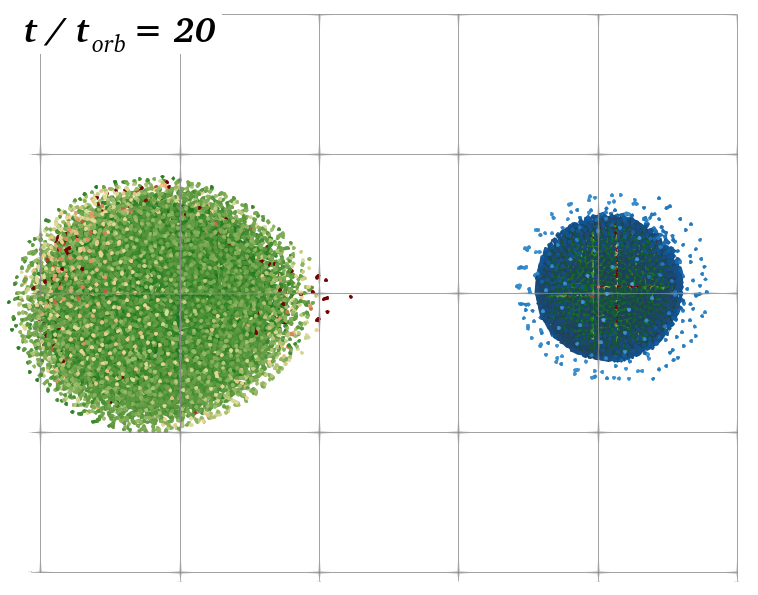} &
  \includegraphics[width=0.34\textwidth]{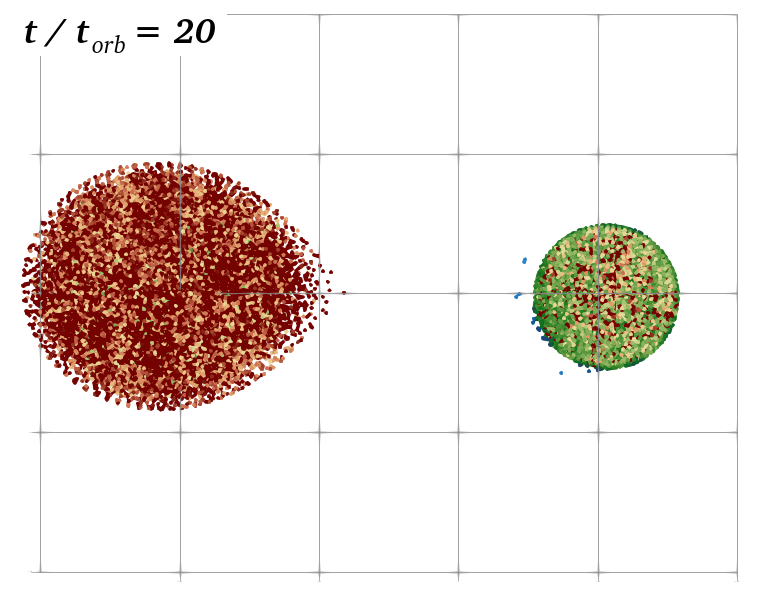}
\end{tabular}
\caption{Evolution of the temperature \edt{in the}{looking down on the orbital plane of the} binary white dwarf system, relaxed using the generic approach (left) vs. the HPR relaxation method (right). 
Four rows, from top to bottom, correspond to four orbital phases: $t/t_{\rm orb} = 0$, 0.1, 1, and 20, where orbital periods are $t_{\rm orb} = 80.88$~s and $79.11$~s for the generic and HPR cases, respectively. 
\edt{Significantly higher temperatures in the envelope are observed for the system relaxed using the generic method: $10^8$~K vs. $\lesssim10^7$~K for the HPR method.
Grid step is $10^9$~cm.}{The difference in the orbital periods is due to the generic binary settling into a slightly wider orbit over the course of the evolution.
By hydrodynamically evolving the particle energies, the temperature is recovered  with the HELMHOLTZ EoS. 
Significantly higher temperatures in the envelope are observed for the system relaxed using the generic method: $10^8$~K vs. $\lesssim10^7$~K for the HPR method. 
The grid step is $10^9$~cm.}}
\label{fig:WDWDbinaryevo}
\end{figure*}

For our generic setup of the binary, we observe that a substantial amount of mass ($10^{-3}M_\odot$) offset from the L1 point has crossed over the Roche lobe for the donor star during the driving phase (see the top left panel in Figure~\ref{fig:WDWDbinaryevo}). 
With this setup, we do not achieve the formation of the Roche cone, which can only be adequately resolved by extending the driving phase over a longer time or introducing a velocity-damping term. 

We found similarities in different binary merger papers suggesting that the systems are driven until mass transfer begins. 
Some groups then turn off the driving force and begin orbital evolution \citep{Motl2017}.
Other groups have noted that the binaries are over-relaxed, i.e., the Roche potential is artificially over-filled at this separation; thus, they modify the binary by increasing the separation \citep{Zenati2019,Bobrick2022}. 
The modification of the orbital separation seems as arbitrary as driving to a predetermined separation, so we choose to drive the system toward the stable separation we found via our relaxation method.
We do not employ velocity damping during driving since this is only mentioned occasionally in the cited literature.

We want to highlight the importance of velocity damping if no other initialization method is implemented. 
The WDB setup prepared with the generic method was driven to the orbital separation of $3.1696\times10^9$~cm. 
Without velocity damping, this configuration distinctly overlaps the critical Roche surface at the start of the complete evolution, forming a broad accretion stream; see Figure~\ref{fig:WDWDbinaryevo}. 

\begin{figure}[ht]
\begin{tabular}{c}
  \includegraphics[width=\columnwidth]{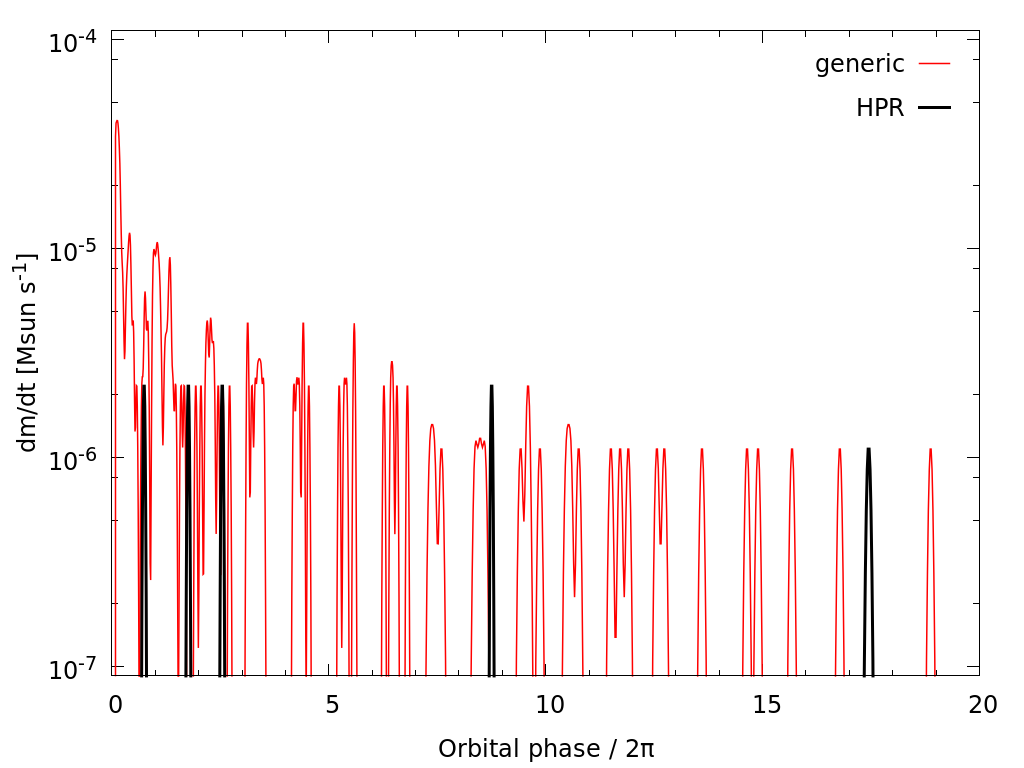}
\end{tabular}
\caption{\edt{The accretion streams formed from the three resolutions taken at 25s of evolution for each binary. The 3D systems are mapped to 1D, where the 1.00 M$_{\odot}$ is on the left of the figure and the 0.50 M$_{\odot}$ donor WD is on the right. From top to bottom: 150k, 200k, and 300k particles.}{The mass transfer over twenty initial orbital periods for the generically relaxed system (red) compared with the HPR relaxed system (black). We find there is significantly less mass transfer over the total time in the HPR system. Additionally, the burst-nature of the mass transfer is evident. The periodic oscillations in the orbital separation and individual breathing modes power the mass accretion through the entire evolution.}
} 
\label{fig:mass_transfer_rate}
\end{figure}

\begin{figure*}[htp!]
\centering
  \includegraphics[width=0.76\textwidth]{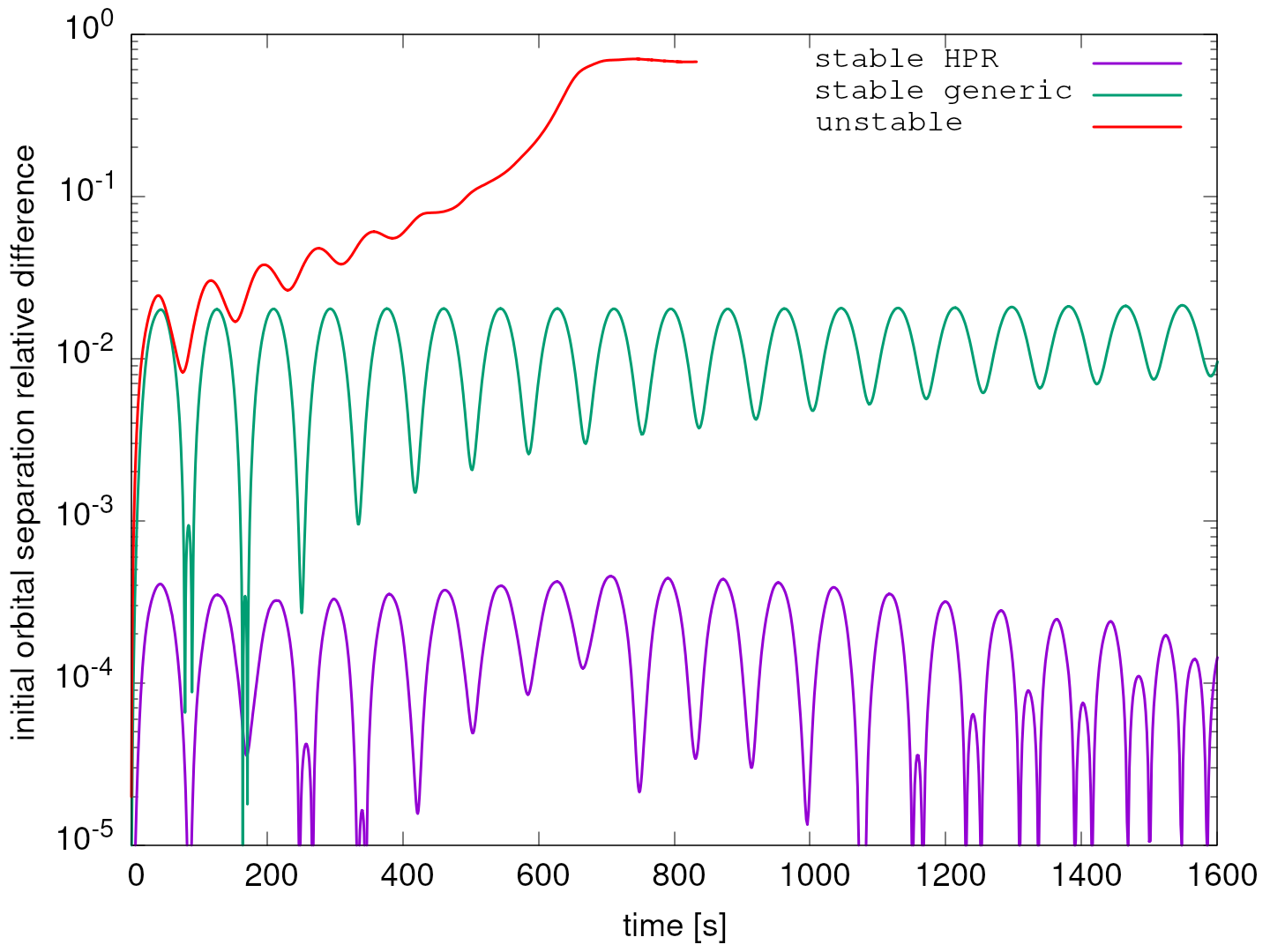}
  \caption{The oscillation modes after mapping orbital velocity at an orbital separation close to the critical distance for the HPR relaxation binary and the general method. The unstable system is relaxed to the critical separation. This unstable binary merges within six orbital periods. The binary relaxed with HPR is stable close to the critical separation for over 20 orbits. The oscillations of the orbit are maximally $4.4\times 10^{-4}$ or 14 km. The binary relaxed to the same separation with a generalized method exhibits oscillations up to $2\times 10^{-2}$ or 640 km. This system also underwent significant periodic mass transfer.}
\label{fig:HPR_general_unstable}
\end{figure*}

After preparing the two configurations, we turn off the driving and proceed with the evolution in the inertial frame. 
Despite the differences between the two setups, we find that mass transfer occurs in both cases. 
However, for the HPR-relaxed binary, only five particles, or $5.1\times 10^{-5} M_\odot$ are transferred over twenty orbits. 
With such a minimal mass transfer, the orbital separation is maintained almost constant, varying with an amplitude of only 14.51~km, or $\sim4\times10^{-4}\ a_{\rm orb}$ (see Figure \ref{fig:HPR_general_unstable}). 
For the binary prepared with the generic relaxation technique, the mass transfer starts strong, as expected. 
However, since the system is over-relaxed, the material that does not immediately transfer is pulled back toward the donor WD. 
This starts a radial breathing mode in the WD, with oscillating radius and orbital separation. 
Due to these oscillations, the mass transfer occurs in periodic bursts, with rates reaching $\sim10^{-5}\ M_\odot\ {\rm s}^{-1}$ per burst, for the first few orbital periods\edt{}{; see Figure~\ref{fig:mass_transfer_rate}}. 

Subsequently, the binary widens, and mass transfer slows down; however, it still occurs at a low rate for the remainder of the simulation. 
A total of $1.6\times 10^{-3} M_\odot$ is transferred over twenty periods for a system that presumably should have remained stable. 
Due to the differences in the dynamical evolution, we find that the temperatures of the WDB can be significantly higher when using the generic relaxation approach versus HPR.
This is shown in Fig.~\ref{fig:WDWDbinaryevo}. 
For example, after twenty periods, the surface temperature of the accreting WD is around $10^8\:$K versus the WD in the HPR-relaxed system whose temperature is $\sim 10^7\:$K. 
Further studies, including nucleosynthesis calculations, are needed to determine whether this difference has consequences on the overall dynamics of the WDB and the composition of the merger.

To compare with our stable case, we proceeded with the driving of the generic case further toward the critical separation.
This case is shown with the solid red line in Figure~\ref{fig:HPR_general_unstable}.
Driving was terminated as soon as the particles transferred to the companion, and the binary was mapped for complete evolution. 
The setup resulted in a runaway mass transfer despite the orbital and radial oscillations evident in the simulations. 
\cite{Dan11} have discussed how these effects have previously overestimated the accretion rates with implications for the final disk, producing higher temperatures and extended disk profiles. 
We find that the binary tidally disrupts over about five orbital periods. 
By driving the systems toward merger, the desired result can be easily obtained; however, the correct physical result may be obscured. 
Accurate initial conditions are of even greater importance for systems where the initial accretion stream is essential.

\section{Conclusion}
\label{sec:conclusion}
In this work, we present a new technique for Smoothed Particle Hydrodynamics (SPH) to efficiently produce accurate equilibrium particle lattices, which can be used to initialize simulations. 
The Halted Pendulum Relaxation (HPR) method, as the name suggests, is based on stopping the particles at the time when the kinetic energy of the system reaches its maximum, similar to stopping a pendulum from oscillating.
We use HPR to create particle distributions for single white dwarfs and white dwarf binaries (WDBs) using the Los Alamos National Laboratory FleCSPH code in combination with an external potential approach to represent gravitational and orbital forces. 

We compare HPR to three other methods for single stars: relaxation in an external potential using velocity damping, relaxation by increased artificial viscosity, and the Weighted Voronoi Tesselation technique.
Our method significantly improves the efficiency and accuracy when setting up SPH particles to replicate a target density profile. 
For binaries, we apply HPR and the Roche lobe potential to set up initial conditions for tidally locked WDBs with mass ratio $q = 0.5$ at (1) a wide orbit, (2) the Roche lobe overflow, and (3) the onset of the accretion stream. 
These conditions are then used as the starting point for a dynamic simulation where we compare the outcome to a WDB system that is relaxed without HPR velocity damping. 
We find significant differences in the mass transfer between both systems, with the HPR-relaxed WDB experiencing a transfer of only a few particles versus the generic-relaxed binary subject to periodic bursts of mass transfer due to the persistence of breathing modes. 
We conclude that for compact star binary merger simulations, especially when the focus is on accurate modeling of material transfer, great care must be taken to damp out numerical high-frequency noise and low-frequency global motion. 
HPR is an effective, straightforward method to implement and apply in existing SPH codes. 

HPR is not limited to compact star modeling but can be applied to any configuration where balancing forces create equilibrium particle configurations. 
Combined with the Fast Multipole Method (FMM) for long-range forces, this can be an efficient approach to \edt{solving}{finding stable equilibrium configurations specified by} elliptic equations in SPH.
In astrophysics, this primarily involves gravitational forces balanced by the material EoS or strength. 
\edt{}{In particular, for the problem studied in this paper, FMM is used to compute gravitational potential without a global elliptic solve. 
This is possible because Poisson equation for the gravitational potential admits solution via a superposition of Green's functions from individual particles sourcing the field.
For more general mechanical system of particles and long-range forces that can be found via a solution of an elliptic equation, an equilibrium particle configuration can be found by HPR, provided that FMM can be used with the Green's functions of this elliptic equation.}
Future potential applications in FleCSPH include simulations in nuclear astrophysics for neutron-star binaries and kilonova modeling, as well as planetary physics in the form of asteroids and impacts.
\section*{Acknowledgment}
This research used resources provided by the Darwin test bed at Los Alamos National Laboratory (LANL), which is funded by the Computational Systems and Software Environments subprogram of LANL's Advanced Simulation and Computing (ASC) program (NNSA/DOE). This work is supported by the LANL ASC program and LANL’s Laboratory Directed Research and Development (LDRD) Program under projects 20200145ER and 20190021DR. M.F. was supported through LANL's Center for Space and Earth Science (CSES). CSES is funded by LANL’s LDRD program under project number 20180475DR. This work used resources provided by the LANL Institutional Computing Program. LANL is operated by Triad National Security, LLC, for the National Nuclear Security Administration of the U.S.DOE (Contract No. 89233218CNA000001). This work is authorized for unlimited release under LA-UR-22-29924.


\bibliographystyle{mnras}
\bibliography{refs}

\begin{thebibliography}{}
\makeatletter
\relax
\def\mn@urlcharsother{\let\do\@makeother \do\$\do\&\do\#\do\^\do\_\do\%\do\~}
\def\mn@doi{\begingroup\mn@urlcharsother \@ifnextchar [ {\mn@doi@}
  {\mn@doi@[]}}
\def\mn@doi@[#1]#2{\def\@tempa{#1}\ifx\@tempa\@empty \href
  {http://dx.doi.org/#2} {doi:#2}\else \href {http://dx.doi.org/#2} {#1}\fi
  \endgroup}
\def\mn@eprint#1#2{\mn@eprint@#1:#2::\@nil}
\def\mn@eprint@arXiv#1{\href {http://arxiv.org/abs/#1} {{\tt arXiv:#1}}}
\def\mn@eprint@dblp#1{\href {http://dblp.uni-trier.de/rec/bibtex/#1.xml}
  {dblp:#1}}
\def\mn@eprint@#1:#2:#3:#4\@nil{\def\@tempa {#1}\def\@tempb {#2}\def\@tempc
  {#3}\ifx \@tempc \@empty \let \@tempc \@tempb \let \@tempb \@tempa \fi \ifx
  \@tempb \@empty \def\@tempb {arXiv}\fi \@ifundefined
  {mn@eprint@\@tempb}{\@tempb:\@tempc}{\expandafter \expandafter \csname
  mn@eprint@\@tempb\endcsname \expandafter{\@tempc}}}

\bibitem[\protect\citeauthoryear{{Arth}, {Donnert}, {Steinwandel}, {B{\"o}ss},
  {Halbesma}, {P{\"u}tz}, {Hubber}  \& {Dolag}}{{Arth} et~al.}{2019}]{Arth2019}
{Arth} A.,  {Donnert} J.,  {Steinwandel} U.,  {B{\"o}ss} L.,  {Halbesma} T.,
  {P{\"u}tz} M.,  {Hubber} D.,   {Dolag} K.,  2019, arXiv e-prints, \href
  {https://ui.adsabs.harvard.edu/abs/2019arXiv190711250A} {p. arXiv:1907.11250}

\bibitem[\protect\citeauthoryear{{Benz}, {Bowers}, {Cameron}  \&
  {Press}}{{Benz} et~al.}{1990}]{Benz1990}
{Benz} W.,  {Bowers} R.~L.,  {Cameron} A.~G.~W.,   {Press} W.~H.~.,  1990,
  \mn@doi [\apj] {10.1086/168273}, \href
  {https://ui.adsabs.harvard.edu/abs/1990ApJ...348..647B} {348, 647}

\bibitem[\protect\citeauthoryear{Bergen \& Charest}{Bergen \&
  Charest}{2017}]{charest2017flexible}
Bergen B.~K.,  Charest M. R.~J.,  2017, Technical report, Flexible
  Computational Science Infrastructure ({FleCSI}): Overview \& Application
  Progress.
Los Alamos National Lab.(LANL), Los Alamos, NM (United States)

\bibitem[\protect\citeauthoryear{{Bobrick}, {Zenati}, {Perets}, {Davies}  \&
  {Church}}{{Bobrick} et~al.}{2022}]{Bobrick2022}
{Bobrick} A.,  {Zenati} Y.,  {Perets} H.~B.,  {Davies} M.~B.,   {Church} R.,
  2022, \mn@doi [\mnras] {10.1093/mnras/stab3574}, \href
  {https://ui.adsabs.harvard.edu/abs/2022MNRAS.510.3758B} {510, 3758}

\bibitem[\protect\citeauthoryear{Chandrasekhar}{Chandrasekhar}{1957}]{Chandrasekhar1957}
Chandrasekhar S.,  1957, An Introduction to the Study of Stellar Structure.
Astrophysical monographs, Dover Publications, \url
  {https://books.google.com/books?id=VwPLAgAAQBAJ}

\bibitem[\protect\citeauthoryear{D'Souza, Motl, Tohline  \& Frank}{D'Souza
  et~al.}{2006}]{DSouza2006}
D'Souza M. C.~R.,  Motl P.~M.,  Tohline J.~E.,   Frank J.,  2006, \mn@doi [The
  Astrophysical Journal] {10.1086/500384}, 643, 381

\bibitem[\protect\citeauthoryear{{Dan}, {Rosswog}  \& {Br{\"u}ggen}}{{Dan}
  et~al.}{2009}]{Dan09}
{Dan} M.,  {Rosswog} S.,   {Br{\"u}ggen} M.,  2009, in Journal of Physics
  Conference Series. p. 012034 (\mn@eprint {arXiv} {0811.1517}),
  \mn@doi{10.1088/1742-6596/172/1/012034}

\bibitem[\protect\citeauthoryear{{Dan}, {Rosswog}, {Guillochon}  \&
  {Ramirez-Ruiz}}{{Dan} et~al.}{2011}]{Dan11}
{Dan} M.,  {Rosswog} S.,  {Guillochon} J.,   {Ramirez-Ruiz} E.,  2011, \mn@doi
  [\apj] {10.1088/0004-637X/737/2/89}, \href
  {https://ui.adsabs.harvard.edu/abs/2011ApJ...737...89D} {737, 89}

\bibitem[\protect\citeauthoryear{{Dan}, {Rosswog}, {Guillochon}  \&
  {Ramirez-Ruiz}}{{Dan} et~al.}{2012}]{Dan12}
{Dan} M.,  {Rosswog} S.,  {Guillochon} J.,   {Ramirez-Ruiz} E.,  2012, \mn@doi
  [\mnras] {10.1111/j.1365-2966.2012.20794.x}, \href
  {https://ui.adsabs.harvard.edu/abs/2012MNRAS.422.2417D} {422, 2417}

\bibitem[\protect\citeauthoryear{{Dan}, {Rosswog}, {Br{\"u}ggen}  \&
  {Podsiadlowski}}{{Dan} et~al.}{2014}]{Dan2014}
{Dan} M.,  {Rosswog} S.,  {Br{\"u}ggen} M.,   {Podsiadlowski} P.,  2014,
  \mn@doi [\mnras] {10.1093/mnras/stt1766}, \href
  {https://ui.adsabs.harvard.edu/abs/2014MNRAS.438...14D} {438, 14}

\bibitem[\protect\citeauthoryear{Dehnen}{Dehnen}{2002}]{Dehnen2002}
Dehnen W.,  2002, \mn@doi [Journal of Computational Physics]
  {https://doi.org/10.1006/jcph.2002.7026}, 179, 27

\bibitem[\protect\citeauthoryear{{Dehnen}}{{Dehnen}}{2014}]{Dehnen2014}
{Dehnen} W.,  2014, \mn@doi [Computational Astrophysics and Cosmology]
  {10.1186/s40668-014-0001-7}, \href
  {https://ui.adsabs.harvard.edu/abs/2014ComAC...1....1D} {1, 1}

\bibitem[\protect\citeauthoryear{Diehl, Rockefeller, Fryer, Riethmiller  \&
  Statler}{Diehl et~al.}{2015}]{Diehl2015}
Diehl S.,  Rockefeller G.,  Fryer C.~L.,  Riethmiller D.,   Statler T.~S.,
  2015, \mn@doi [Publications of the Astronomical Society of Australia]
  {10.1017/pasa.2015.50}, 32, e048

\bibitem[\protect\citeauthoryear{{Eggleton}}{{Eggleton}}{1983}]{Eggleton1983}
{Eggleton} P.~P.,  1983, \mn@doi [\apj] {10.1086/160960}, \href
  {https://ui.adsabs.harvard.edu/abs/1983ApJ...268..368E} {268, 368}

\bibitem[\protect\citeauthoryear{{Ellinger}, {Young}, {Fryer}  \&
  {Rockefeller}}{{Ellinger} et~al.}{2012}]{Ellinger2012}
{Ellinger} C.~I.,  {Young} P.~A.,  {Fryer} C.~L.,   {Rockefeller} G.,  2012,
  \mn@doi [\apj] {10.1088/0004-637X/755/2/160}, 755, 160

\bibitem[\protect\citeauthoryear{{Emden}}{{Emden}}{1907}]{Emden1907}
{Emden} R.,  1907, {Gaskugeln}

\bibitem[\protect\citeauthoryear{Even \& Tohline}{Even \&
  Tohline}{2009}]{Even2009}
Even W.,  Tohline J.~E.,  2009, \mn@doi [The Astrophysical Journal Supplement
  Series] {10.1088/0067-0049/184/2/248}, 184, 248

\bibitem[\protect\citeauthoryear{{Fryer} \& {Diehl}}{{Fryer} \&
  {Diehl}}{2008}]{Fryer2008}
{Fryer} C.~L.,  {Diehl} S.,  2008, in {Werner} A.,  {Rauch} T.,  eds,
  Astronomical Society of the Pacific Conference Series Vol. 391,
  Hydrogen-Deficient Stars. p.~335 (\mn@eprint {arXiv} {0711.0864})

\bibitem[\protect\citeauthoryear{{Fryer}, {Rockefeller}  \& {Warren}}{{Fryer}
  et~al.}{2006}]{Fryer2006}
{Fryer} C.~L.,  {Rockefeller} G.,   {Warren} M.~S.,  2006, \mn@doi [\apj]
  {10.1086/501493}, \href
  {https://ui.adsabs.harvard.edu/abs/2006ApJ...643..292F} {643, 292}

\bibitem[\protect\citeauthoryear{{Fryer} et~al.,}{{Fryer}
  et~al.}{2010}]{Fryer2010}
{Fryer} C.~L.,  et~al., 2010, \mn@doi [\apj] {10.1088/0004-637X/725/1/296},
  \href {https://ui.adsabs.harvard.edu/abs/2010ApJ...725..296F} {725, 296}

\bibitem[\protect\citeauthoryear{{Gingold} \& {Monaghan}}{{Gingold} \&
  {Monaghan}}{1977}]{Gingold1977}
{Gingold} R.~A.,  {Monaghan} J.~J.,  1977, \mn@doi [\mnras]
  {10.1093/mnras/181.3.375}, \href
  {https://ui.adsabs.harvard.edu/abs/1977MNRAS.181..375G} {181, 375}

\bibitem[\protect\citeauthoryear{{Guerrero}, {Garc{\'\i}a-Berro}  \&
  {Isern}}{{Guerrero} et~al.}{2004}]{Guerrero2004}
{Guerrero} J.,  {Garc{\'\i}a-Berro} E.,   {Isern} J.,  2004, \mn@doi [\aap]
  {10.1051/0004-6361:20031504}, \href
  {https://ui.adsabs.harvard.edu/abs/2004A&A...413..257G} {413, 257}

\bibitem[\protect\citeauthoryear{{Hachisu}}{{Hachisu}}{1986a}]{Hachisu1986a}
{Hachisu} I.,  1986a, \mn@doi [\apjs] {10.1086/191121}, \href
  {https://ui.adsabs.harvard.edu/abs/1986ApJS...61..479H} {61, 479}

\bibitem[\protect\citeauthoryear{{Hachisu}}{{Hachisu}}{1986b}]{Hachisu1986b}
{Hachisu} I.,  1986b, \mn@doi [\apjs] {10.1086/191148}, \href
  {https://ui.adsabs.harvard.edu/abs/1986ApJS...62..461H} {62, 461}

\bibitem[\protect\citeauthoryear{{Hopkins}}{{Hopkins}}{2013}]{Hopkins2013}
{Hopkins} P.~F.,  2013, \mn@doi [\mnras] {10.1093/mnras/sts210}, \href
  {https://ui.adsabs.harvard.edu/abs/2013MNRAS.428.2840H} {428, 2840}

\bibitem[\protect\citeauthoryear{{Hoyle} \& {Fowler}}{{Hoyle} \&
  {Fowler}}{1960}]{Hoyle1960}
{Hoyle} F.,  {Fowler} W.~A.,  1960, \mn@doi [\apj] {10.1086/146963}, \href
  {https://ui.adsabs.harvard.edu/abs/1960ApJ...132..565H} {132, 565}

\bibitem[\protect\citeauthoryear{{J. Loiseau} \& {et al.}}{{J. Loiseau} \& {et
  al.}}{2020}]{Loiseau2020a}
{J. Loiseau} {et al.} 2020, \mn@doi [SoftwareX]
  {https://doi.org/10.1016/j.softx.2020.100602}, 12, 100602

\bibitem[\protect\citeauthoryear{{Kadam}, {Motl}, {Marcello}, {Frank}  \&
  {Clayton}}{{Kadam} et~al.}{2018}]{Kadam2018}
{Kadam} K.,  {Motl} P.~M.,  {Marcello} D.~C.,  {Frank} J.,   {Clayton} G.~C.,
  2018, \mn@doi [\mnras] {10.1093/mnras/sty2540}, \href
  {https://ui.adsabs.harvard.edu/abs/2018MNRAS.481.3683K} {481, 3683}

\bibitem[\protect\citeauthoryear{{Katz}, {Zingale}, {Calder}, {Swesty},
  {Almgren}  \& {Zhang}}{{Katz} et~al.}{2016}]{Katz2016}
{Katz} M.~P.,  {Zingale} M.,  {Calder} A.~C.,  {Swesty} F.~D.,  {Almgren}
  A.~S.,   {Zhang} W.,  2016, \mn@doi [\apj] {10.3847/0004-637X/819/2/94},
  \href {https://ui.adsabs.harvard.edu/abs/2016ApJ...819...94K} {819, 94}

\bibitem[\protect\citeauthoryear{{Kegerreis} et~al.,}{{Kegerreis}
  et~al.}{2018}]{Kegerreis2018}
{Kegerreis} J.~A.,  et~al., 2018, \mn@doi [\apj] {10.3847/1538-4357/aac725},
  861, 52

\bibitem[\protect\citeauthoryear{{Korobkin}, {Lim}, {Sagert}, {Loiseau},
  {Mauney}, {Kaltenborn}, {Tsao}  \& {Even}}{{Korobkin}
  et~al.}{2021}]{Korobkin2021}
{Korobkin} O.,  {Lim} H.,  {Sagert} I.,  {Loiseau} J.,  {Mauney} C.,
  {Kaltenborn} M. A.~R.,  {Tsao} B.-J.,   {Even} W.~P.,  2021, arXiv e-prints,
  \href {https://ui.adsabs.harvard.edu/abs/2021arXiv210707166K} {p.
  arXiv:2107.07166}

\bibitem[\protect\citeauthoryear{Lane \& Lane}{Lane \& Lane}{1870}]{Lane1870}
Lane H.~J.,  Lane J.~H.,  1870, {On the theoretical temperature of the Sun,
  under the hypothesis of a gaseous mass maintaining its volume by its internal
  heat, and depending on the laws of gases as known to terrestrial experiment},
  \mn@doi{10.2475/ajs.s2-50.148.57}, \url
  {https://doi.org/10.2475/ajs.s2-50.148.57}

\bibitem[\protect\citeauthoryear{{Loiseau}, {Lim}, {Kaltenborn}, {Korobkin},
  {Mauney}, {Sagert}, {Even}  \& {Bergen}}{{Loiseau}
  et~al.}{2020}]{Loiseau2020b}
{Loiseau} J.,  {Lim} H.,  {Kaltenborn} M.~A.,  {Korobkin} O.,  {Mauney} C.~M.,
  {Sagert} I.,  {Even} W.~P.,   {Bergen} B.~K.,  2020, {FleCSPH: Parallel and
  distributed SPH implementation based on the FleCSI}, Astrophysics Source Code
  Library, record ascl:2007.011 (\mn@eprint {ascl} {2007.011})

\bibitem[\protect\citeauthoryear{{Lombardi}, {Sills}, {Rasio}  \&
  {Shapiro}}{{Lombardi} et~al.}{1999}]{Lombardi1999}
{Lombardi} J.~C.,  {Sills} A.,  {Rasio} F.~A.,   {Shapiro} S.~L.,  1999,
  \mn@doi [Journal of Computational Physics] {10.1006/jcph.1999.6256}, \href
  {https://ui.adsabs.harvard.edu/abs/1999JCoPh.152..687L} {152, 687}

\bibitem[\protect\citeauthoryear{{Maindl}, {Sch{\"a}fer}, {Speith}, {S{\"u}li},
  {Forg{\'a}cs-Dajka}  \& {Dvorak}}{{Maindl} et~al.}{2013}]{Maindl2013}
{Maindl} T.~I.,  {Sch{\"a}fer} C.,  {Speith} R.,  {S{\"u}li} {\'A}.,
  {Forg{\'a}cs-Dajka} E.,   {Dvorak} R.,  2013, \mn@doi [Astronomische
  Nachrichten] {10.1002/asna.201311979}, 334, 996

\bibitem[\protect\citeauthoryear{{Marcello}}{{Marcello}}{2017}]{Marcello2017}
{Marcello} D.~C.,  2017, \mn@doi [Astrophysical Journal]
  {10.3847/1538-3881/aa7b2f}, \href
  {https://ui.adsabs.harvard.edu/abs/2017AJ....154...92M} {154, 92}

\bibitem[\protect\citeauthoryear{{Marcello}, {Shiber}, {De Marco}, {Frank},
  {Clayton}, {Motl}, {Diehl}  \& {Kaiser}}{{Marcello}
  et~al.}{2021}]{Marcello2021}
{Marcello} D.~C.,  {Shiber} S.,  {De Marco} O.,  {Frank} J.,  {Clayton} G.~C.,
  {Motl} P.~M.,  {Diehl} P.,   {Kaiser} H.,  2021, \mn@doi [\mnras]
  {10.1093/mnras/stab937}, \href
  {https://ui.adsabs.harvard.edu/abs/2021MNRAS.504.5345M} {504, 5345}

\bibitem[\protect\citeauthoryear{Monaghan}{Monaghan}{1992}]{Monaghan1992}
Monaghan J.~J.,  1992, \mn@doi [Annual Review of Astronomy and Astrophysics]
  {10.1146/annurev.aa.30.090192.002551}, 30, 543

\bibitem[\protect\citeauthoryear{{Motl}, {Tohline}  \& {Frank}}{{Motl}
  et~al.}{2002}]{Motl2002}
{Motl} P.~M.,  {Tohline} J.~E.,   {Frank} J.,  2002, \mn@doi [\apjs]
  {10.1086/324159}, \href
  {https://ui.adsabs.harvard.edu/abs/2002ApJS..138..121M} {138, 121}

\bibitem[\protect\citeauthoryear{{Motl}, {Frank}, {Tohline}  \&
  {D'Souza}}{{Motl} et~al.}{2007}]{Motl2007}
{Motl} P.~M.,  {Frank} J.,  {Tohline} J.~E.,   {D'Souza} M. C.~R.,  2007,
  \mn@doi [\apj] {10.1086/522076}, \href
  {https://ui.adsabs.harvard.edu/abs/2007ApJ...670.1314M} {670, 1314}

\bibitem[\protect\citeauthoryear{{Motl}, {Frank}, {Staff}, {Clayton}, {Fryer},
  {Even}, {Diehl}  \& {Tohline}}{{Motl} et~al.}{2017}]{Motl2017}
{Motl} P.~M.,  {Frank} J.,  {Staff} J.,  {Clayton} G.~C.,  {Fryer} C.~L.,
  {Even} W.,  {Diehl} S.,   {Tohline} J.~E.,  2017, \mn@doi [\apjs]
  {10.3847/1538-4365/aa5bde}, \href
  {https://ui.adsabs.harvard.edu/abs/2017ApJS..229...27M} {229, 27}

\bibitem[\protect\citeauthoryear{{Owen}}{{Owen}}{2012}]{spheral}
{Owen} M.,  2012, SPHERAL, \url {https://spheral.readthedocs.io/en/latest/}

\bibitem[\protect\citeauthoryear{{Pakmor}, {Springel}, {Bauer}, {Mocz},
  {Munoz}, {Ohlmann}, {Schaal}  \& {Zhu}}{{Pakmor} et~al.}{2016}]{Pakmor2016}
{Pakmor} R.,  {Springel} V.,  {Bauer} A.,  {Mocz} P.,  {Munoz} D.~J.,
  {Ohlmann} S.~T.,  {Schaal} K.,   {Zhu} C.,  2016, \mn@doi [\mnras]
  {10.1093/mnras/stv2380}, \href
  {https://ui.adsabs.harvard.edu/abs/2016MNRAS.455.1134P} {455, 1134}

\bibitem[\protect\citeauthoryear{{Price} \& et al.}{{Price} \&
  et~al.}{2018}]{Price2018}
{Price} D.~J.,  et al. 2018, \mn@doi [Publications of the Astronomical Society
  of Australia] {10.1017/pasa.2018.25}, \href
  {https://ui.adsabs.harvard.edu/abs/2018PASA...35...31P} {35, e031}

\bibitem[\protect\citeauthoryear{{Rasio} \& {Shapiro}}{{Rasio} \&
  {Shapiro}}{1995}]{Rasio1995}
{Rasio} F.~A.,  {Shapiro} S.~L.,  1995, \mn@doi [\apj] {10.1086/175130}, \href
  {https://ui.adsabs.harvard.edu/abs/1995ApJ...438..887R} {438, 887}

\bibitem[\protect\citeauthoryear{{Raskin}, {Scannapieco}, {Fryer},
  {Rockefeller}  \& {Timmes}}{{Raskin} et~al.}{2012}]{Raskin2012}
{Raskin} C.,  {Scannapieco} E.,  {Fryer} C.,  {Rockefeller} G.,   {Timmes}
  F.~X.,  2012, \mn@doi [\apj] {10.1088/0004-637X/746/1/62}, \href
  {https://ui.adsabs.harvard.edu/abs/2012ApJ...746...62R} {746, 62}

\bibitem[\protect\citeauthoryear{Raskin, Kasen, Moll, Schwab  \&
  Woosley}{Raskin et~al.}{2014}]{Raskin2014}
Raskin C.,  Kasen D.,  Moll R.,  Schwab J.,   Woosley S.,  2014, \mn@doi [The
  Astrophysical Journal] {10.1088/0004-637x/788/1/75}, 788, 75

\bibitem[\protect\citeauthoryear{{Rosswog}}{{Rosswog}}{2015}]{Rosswog2015}
{Rosswog} S.,  2015, \mn@doi [\mnras] {10.1093/mnras/stv225}, \href
  {https://ui.adsabs.harvard.edu/abs/2015MNRAS.448.3628R} {448, 3628}

\bibitem[\protect\citeauthoryear{Rosswog}{Rosswog}{2020}]{Rosswog2020}
Rosswog S.,  2020, \mn@doi [Monthly Notices of the Royal Astronomical Society]
  {10.1093/mnras/staa2591}, 498, 4230

\bibitem[\protect\citeauthoryear{{Rosswog} \& {Diener}}{{Rosswog} \&
  {Diener}}{2021}]{Rosswog2021}
{Rosswog} S.,  {Diener} P.,  2021, \mn@doi [Classical and Quantum Gravity]
  {10.1088/1361-6382/abee65}, \href
  {https://ui.adsabs.harvard.edu/abs/2021CQGra..38k5002R} {38, 115002}

\bibitem[\protect\citeauthoryear{{Rosswog}, {Liebend{\"o}rfer}, {Thielemann},
  {Davies}, {Benz}  \& {Piran}}{{Rosswog} et~al.}{1999}]{Rosswog1999}
{Rosswog} S.,  {Liebend{\"o}rfer} M.,  {Thielemann} F.~K.,  {Davies} M.~B.,
  {Benz} W.,   {Piran} T.,  1999, \aap, 341, 499

\bibitem[\protect\citeauthoryear{{Sagert}, {Korobkin}, {Tews}, {Tsao}, {Lim},
  {Falato}  \& {Loiseau}}{{Sagert} et~al.}{2022}]{Sagert2022}
{Sagert} I.,  {Korobkin} O.,  {Tews} I.,  {Tsao} B.-J.,  {Lim} H.,  {Falato}
  M.~J.,   {Loiseau} J.,  2022, arXiv e-prints, \href
  {https://ui.adsabs.harvard.edu/abs/2022arXiv221101927S} {p. arXiv:2211.01927}

\bibitem[\protect\citeauthoryear{{Saitoh} \& {Makino}}{{Saitoh} \&
  {Makino}}{2013}]{Saitoh2013}
{Saitoh} T.~R.,  {Makino} J.,  2013, \mn@doi [\apj]
  {10.1088/0004-637X/768/1/44}, \href
  {https://ui.adsabs.harvard.edu/abs/2013ApJ...768...44S} {768, 44}

\bibitem[\protect\citeauthoryear{{Schaefer}, {Riecker}, {Wandel}, {Maindl},
  {Scherrer}, {Werner}, {Burger}  \& {Morlock}}{{Schaefer}
  et~al.}{2019}]{Schaefer2019}
{Schaefer} C.~M.,  {Riecker} S.,  {Wandel} O.,  {Maindl} T.~I.,  {Scherrer} S.,
   {Werner} J.,  {Burger} C.,   {Morlock} M.,  2019, {miluphcuda: Smooth
  particle hydrodynamics code}, Astrophysics Source Code Library, record
  ascl:1911.023 (\mn@eprint {ascl} {1911.023})

\bibitem[\protect\citeauthoryear{{Segretain}, {Chabrier}  \&
  {Mochkovitch}}{{Segretain} et~al.}{1997}]{Segretain1997}
{Segretain} L.,  {Chabrier} G.,   {Mochkovitch} R.,  1997, \mn@doi [\apj]
  {10.1086/304015}, \href
  {https://ui.adsabs.harvard.edu/abs/1997ApJ...481..355S} {481, 355}

\bibitem[\protect\citeauthoryear{{Shariff}, {Jiao}, {Trotta}  \& {van
  Dyk}}{{Shariff} et~al.}{2016}]{Shariff2016}
{Shariff} H.,  {Jiao} X.,  {Trotta} R.,   {van Dyk} D.~A.,  2016, \mn@doi
  [\apj] {10.3847/0004-637X/827/1/1}, \href
  {https://ui.adsabs.harvard.edu/abs/2016ApJ...827....1S} {827, 1}

\bibitem[\protect\citeauthoryear{Stewart et~al.,}{Stewart
  et~al.}{2022}]{Stewart2022}
Stewart A.~R.,  et~al., 2022, in Scientific Visualization Supercomputing
  Conference 2021, St. Louis, MO.

\bibitem[\protect\citeauthoryear{{Tegmark}}{{Tegmark}}{1996}]{Tegmark96}
{Tegmark} M.,  1996, \mn@doi [\apjl] {10.1086/310310}, \href
  {https://ui.adsabs.harvard.edu/abs/1996ApJ...470L..81T} {470, L81}

\bibitem[\protect\citeauthoryear{{Timmes} \& {Swesty}}{{Timmes} \&
  {Swesty}}{2000}]{Timmes2000}
{Timmes} F.~X.,  {Swesty} F.~D.,  2000, \mn@doi [\apjs] {10.1086/313304}, \href
  {https://ui.adsabs.harvard.edu/abs/2000ApJS..126..501T} {126, 501}

\bibitem[\protect\citeauthoryear{{Tsao et al.}}{{Tsao et al.}}{2021}]{Tsao2021}
{Tsao et al.} B.-J.,  2021, in Proc. of 2021 International SPHERIC workshop
  (virtual). New Jersey Institute of Technology

\bibitem[\protect\citeauthoryear{{Wang} \& {White}}{{Wang} \&
  {White}}{2007}]{Wang2007}
{Wang} J.,  {White} S. D.~M.,  2007, \mn@doi [\mnras]
  {10.1111/j.1365-2966.2007.12053.x}, \href
  {https://ui.adsabs.harvard.edu/abs/2007MNRAS.380...93W} {380, 93}

\bibitem[\protect\citeauthoryear{{White}}{{White}}{1996}]{White1996}
{White} S.~D.~M.,  1996, in {Schaeffer} R.,  {Silk} J.,  {Spiro} M.,
  {Zinn-Justin} J.,  eds, Cosmology and Large Scale Structure. p.~349

\bibitem[\protect\citeauthoryear{{Yoon}, {Podsiadlowski}  \& {Rosswog}}{{Yoon}
  et~al.}{2007}]{Yoon2007}
{Yoon} S.~C.,  {Podsiadlowski} P.,   {Rosswog} S.,  2007, \mn@doi [\mnras]
  {10.1111/j.1365-2966.2007.12161.x}, \href
  {https://ui.adsabs.harvard.edu/abs/2007MNRAS.380..933Y} {380, 933}

\bibitem[\protect\citeauthoryear{{Yoshida}}{{Yoshida}}{2019}]{Yoshida2019}
{Yoshida} S.,  2019, \mn@doi [\mnras] {10.1093/mnras/stz1030}, \href
  {https://ui.adsabs.harvard.edu/abs/2019MNRAS.486.2982Y} {486, 2982}

\bibitem[\protect\citeauthoryear{{Yungelson} \& {Kuranov}}{{Yungelson} \&
  {Kuranov}}{2017}]{Yungelson2017}
{Yungelson} L.~R.,  {Kuranov} A.~G.,  2017, \mn@doi [\mnras]
  {10.1093/mnras/stw2432}, \href
  {https://ui.adsabs.harvard.edu/abs/2017MNRAS.464.1607Y} {464, 1607}

\bibitem[\protect\citeauthoryear{{Zenati}, {Perets}  \& {Toonen}}{{Zenati}
  et~al.}{2019}]{Zenati2019}
{Zenati} Y.,  {Perets} H.~B.,   {Toonen} S.,  2019, \mn@doi [\mnras]
  {10.1093/mnras/stz316}, \href
  {https://ui.adsabs.harvard.edu/abs/2019MNRAS.486.1805Z} {486, 1805}

\bibitem[\protect\citeauthoryear{{Zhu}, {Chang}, {van Kerkwijk}  \&
  {Wadsley}}{{Zhu} et~al.}{2013}]{Zhu2013}
{Zhu} C.,  {Chang} P.,  {van Kerkwijk} M.~H.,   {Wadsley} J.,  2013, \mn@doi
  [\apj] {10.1088/0004-637X/767/2/164}, \href
  {https://ui.adsabs.harvard.edu/abs/2013ApJ...767..164Z} {767, 164}

\bibitem[\protect\citeauthoryear{{van Kerkwijk}, {Chang}  \& {Justham}}{{van
  Kerkwijk} et~al.}{2010}]{vanKerkwijk2010}
{van Kerkwijk} M.~H.,  {Chang} P.,   {Justham} S.,  2010, \mn@doi [\apjl]
  {10.1088/2041-8205/722/2/L157}, \href
  {https://ui.adsabs.harvard.edu/abs/2010ApJ...722L.157V} {722, L157}

\makeatother
\end{thebibliography}



\end{document}